\documentclass[aps,twocolumn,prc,superscriptaddress,showpacs,nofootinbib,floatfix,amssymb,amsfonts,amsmath]{revtex4-1}

\usepackage{graphicx}
\usepackage{dcolumn}
\usepackage{bm}
\usepackage{xcolor}
\usepackage{amsmath}    
\usepackage{amsfonts}   
\usepackage{amssymb}
\usepackage{graphicx}   
\usepackage{multirow} 

\begin{document}

\title{Global optimization of harmonic oscillator basis in covariant density functional theory}

\author{B. Osei}
\affiliation{Department of Physics and Astronomy, Mississippi
State University, MS 39762}

\author{A. V. Afanasjev}
\affiliation{Department of Physics and Astronomy, Mississippi
State University, MS 39762}

\author{A. Dalbah}
\affiliation{Department of Physics and Astronomy, Mississippi
State University, MS 39762}

\date{\today}

\begin{abstract}

The present investigation focuses on the improvement of the accuracy
of the description of binding energies within moderately sized fermionic basis.
Using the solutions corresponding to infinite fermionic basis it was shown 
that in the case of meson exchange (ME)  covariant energy density 
functionals (CEDFs)  the global accuracy of the description of binding energies in the finite 
$N_F=16-20$  bases can be drastically (by a factor ranging from $\approx 3$ up
to $\approx 9$ dependent on the functional and $N_F$)  improved by 
a global  optimization of oscillator frequency of the basis.  This is a consequence 
of  the unique  feature of the ME functionals in which with increasing fermionic basis size fermionic 
and mesonic energies approach  the exact (infinite basis) solution from above 
and below,  respectively. As a consequence, an optimal 
oscillator frequency 
$\hbar\omega_0$  of the basis can be defined which provides an accurate reproduction 
of exact total binding energies by the ones calculated in truncated basis. This leads to a very 
high accuracy of  the calculations in moderately sized $N_F=20$ basis when 
mass dependent oscillator frequency is used:  global rms differences  
$\delta B_{rms}$ between the binding energies calculated in infinite and truncated bases 
are only 0.025  MeV and 0.031 MeV for the  NL5(Z) and DD-MEZ functionals, respectively.  
Optimized values of the oscillator  frequency $\hbar\omega_0$ are provided for three major 
classes  of CEDFs, i.e. for density dependent meson exchange functionals,  nonlinear meson 
exchange ones and point coupling functionals.

 \end{abstract}

\maketitle

\section{Introduction}

   The basis set expansion method is a classical method of the solution of  many quantum-mechanical 
problems in different fields such as molecular \cite{HHJKO.99,Lehtola.19} and 
nuclear \cite{Delves-book,CAKKMV.12,TOAPT.24} physics, quantum chemistry
\cite{Lehtola.19},  quantum dots \cite{K.2009}  etc.  Dependent on the type and symmetry of the object 
under study different bases such as harmonic oscillator (HO) 
\cite{GRT.90,NilRag-book,K.2009,CAKKMV.12,TOAPT.24},  Woods-Saxon \cite{ZMR.03,ZPZ.22} and others \cite{HHJKO.99,Lehtola.19} 
are  used for the calculation of its properties. In most of the applications the basis is truncated due 
to numerical limitations. In such a situation,  two major  questions emerge. First, how accurate is the 
description of physical observables in truncated basis as compared with exact solution? In many cases, 
the answer on such a question does not exist because of the absence of numerically accurate exact solution 
corresponding to infinite basis\footnote{For example, the assessment of the accuracy of the truncation of the HO basis 
has been either not carried out or only performed by comparing the solutions obtained with $N_F$ and $N_{F + 2}$ full 
fermionic  shells
in the CDFT publications
(see Sec. V of Ref.\ \cite{TOAPT.24} for a short review). It is only in Refs.\ \cite{TOAPT.24,OATDPDF.25} that such an 
assessment has been done with respect of infinite basis solutions in the CDFT.
 Similar 
situation exists also in many non-relativistic DFT calculations (see, for example, fitting protocols of 
the D1 \cite{D1} and D1S \cite{D1S} Gogny forces and the UNEDF* family of the Skyrme forces 
\cite{UNEDF0,UNEDF2}).}.  
Second, what is a 
convergence rate for a given physical observable as a function of the size and parameters of 
the basis  and whether  it is smooth enough to generate effective extrapolation 
procedure to infinite basis? This rate depends on different factors such as  the type of interaction 
model  (chemical potential, nuclear potential, two-body interaction, meson-nucleon coupling etc),  
specific system being considered (nuclei with different values of $Z$ and $N$, molecules, quantum dots 
etc) and  technical details of numerical calculations (see Refs.\ 
\cite{Delves-book,HHJKO.99,K.2009,CAKKMV.12,Lehtola.19}).

The HO basis in widely used in the nuclear physics applications because of its simplicity 
(see Refs.\ \cite{D1S,GRT.90,NilRag-book,UNEDF0,CAKKMV.12}).  While the 
investigations of the extrapolation features of the solutions based on HO from small to very large 
(basically infinite) basis have been in the focus of effective field and {\it ab initio} 
communities in recent years (see Refs.\ \cite{CAKKMV.12,FHP.12,MEFHP.13,BEHPW.16,CK.16}), 
such efforts were very limited  in the  framework of covariant density functional theory (CDFT).

  The CDFT with meson exchange functionals \cite{Rei.89,VALR.05,RDFNS.16} is unique 
since it has two bases i.e. fermionic and bosonic (mesonic)
because the nucleus is described as a system of nucleons (fermions) which interact via the
exchange of mesons (bosons).  This is contrary to the majority of quantum objects the
description of which requires only one basis. However, the impact of the coupling of these
two bases via respective sectors of the CDFT on the convergence of the solutions has not 
been investigated till now. 

  In more than 90\% of the papers published so far in the CDFT framework, 
 Dirac spinors and the meson fields are 
expanded in terms of HO wave functions with respective symmetries (see Refs.\ 
\cite{DIRHB-code.14,TOAPT.24})  and these expansions include all fermionic and 
bosonic states corresponding to full $N_F$ and $N_B$ fermionic and bosonic 
shells, respectively. However, it is only recently that the extrapolations  to infinite bosonic 
and fermionic bases have been worked out in Refs.\ \cite{TOAPT.24,OATDPDF.25}. 
Moreover, for a limited set of nuclei the numerical solutions corresponding to infinite basis 
have been calculated in extremely large $N_F$ and $N_B$ bases
in Refs.\ \cite{OATDPDF.25} and they allowed to benchmark above mentioned 
extrapolation procedures.

   Based on these results it was recommend to use the bosonic (mesonic) basis with 
$N_B=40$ the solutions in which deviate from the ones corresponding to infinite basis 
by only few keVs (see Ref.\ \cite{OATDPDF.25}).  The use of such basis is important
for the development of new generation of CEDFs which are based on global fits of
experimentally known nuclei. Note that for more than thirty years
the $N_B=20$ basis was standard in the CDFT calculations. However,  in actinides and 
superheavy nuclei the solutions in this basis deviate from infinite basis solutions by up to 
300 and 900 keV in density-dependent meson exchange (DDME) and non-linear meson 
exchange (NLME) covariant energy density functionals (CEDFs), respectively (see 
Ref.\ \cite{OATDPDF.25}). The computational cost of the increase of $N_B$ from 20
to 40 is small.

    However, the situation is much more complicated in the fermionic sector of the 
CDFT: the computational cost increases by approximately  two orders of magnitude 
on transition from $N_F=20$ to $N_F=40$ (see Ref.\ \cite{OATDPDF.25}). This transition is 
also associated with drastic increase of required  memory.  Although manageable on existing
computers the extrapolation procedures from finite to infinite fermionic basis 
suggested in Refs.\ \cite{TOAPT.24,OATDPDF.25} are still numerically expensive. 

   Thus, the major goal of the present paper is the search for alternative methods which 
will allow substantial reduction of the difference between infinite basis results and those obtained 
in finite $N_F$ one keeping the size of $N_F$ moderate and manageable at global 
scale with available computers.  The basic idea of our approach is global optimization
of the HO basis. From 1990 the oscillator frequency of the  HO basis has been fixed at
$\hbar \omega_0 = 41 A^{-1/3}$  [MeV] in existing CDFT calculations 
\cite{GRT.90,AKR.96,RGL.97,DIRHB-code.14}. However, this value has been defined 
from the analysis of only spherical $^{16}$O and $^{208}$Pb nuclei with the NL1 functional 
(see Ref.\ \cite{GRT.90}). Thus, in the present paper the HO basis is specified by a more 
general expression
\begin{eqnarray}
\hbar \omega_0 = f \times  41 A^{-1/3} \,\, {\rm [MeV]}
\end{eqnarray} 
where $f$ is the scaling factor the value of which is defined from a global comparison
of the results obtained in the infinite and finite (truncated at $N_F$) bases. 
Two versions of scaling factor $f$ (globally fixed and mass dependent) are studied
in the present paper. Note that in this
comparison we focus on binding energies which can be extremely precisely defined
in experiment (see Introduction to Ref.\ \cite{TOAPT.24}). Alternative observables 
(such as radii and deformations) are typically measured with higher uncertainties 
which exceed the calculated errors defined from  the comparison of  the $N_F=20$ 
and infinite basis results.

   In addition, the similarities and differences in the convergence pattern of binding 
energies as a function of fermionic basis size as well as their microscopic sources have been 
investigated for three major classes of CEDFs, i.e. for density dependent meson exchange 
(DDME), nonlinear meson exchange (NLME) and point coupling (PC) functionals.

 The paper is organized as follows. Theoretical framework is discussed in Sec.\ \ref{theory}.
  Sec.\ \ref{Quadratures} considers the dependence of the 
results on the number of integration points in the Gauss-Hermite and Gauss-Laguerre quadratures
and the impact of the size of  fermionic basis. The convergence of  nuclear binding energies 
as a function of size of fermionic basis is analyzed  for major classes of the functionals on
selected set of spherical and deformed nuclei in Sec.\ \ref{conv-samples}.  Sec.\ \ref{variat-omega} 
discusses the usefulness of oscillator frequency $\hbar\omega_0$ as a variational parameter.
The impact of coupling of fermionic and bosonic bases via respective sectors of the CDFT on 
convergence of binding energies is examined in Sec.\ \ref{Interconnection}. Global analysis of 
convergence errors for moderately sized  fermionic bases is presented in Sec.\ \ref{global-convergence}. 
Sec.\ \ref{HO-ME-opt} discusses the  optimization of the HO basis for meson exchange functionals.  
The mass dependence of oscillator frequency of the HO basis is considered in Sec.\ \ref{Mass-dep}.
Finally,  Sec.\ \ref{Concl} summarizes the  results  of our paper.

\section{Theoretical framework and the details of the calculations}
\label{theory}

\begin{figure}[htb]
 \begin{center}
   \centering
   \includegraphics*[width=8.5cm]{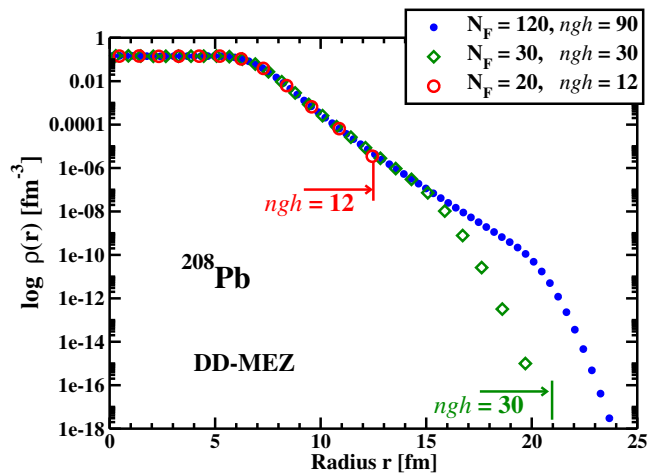}
   \caption{ Neutron density of $^{208}$Pb at the GH integration points for 
   indicated calculational schemes. Lines and arrows indicate the position
   of the last GH integration point i.e. approximate extension of the nucleus
   covered by the calculations. 
    The calculations are carried out with scaling factor $f=1.4$.
 \label{208Pb-density} 
}
\end{center}
\end{figure}

     The numerical calculations are performed in the framework of relativistic 
Hartree-Boboliubov  (RHB) theory using spherical and axially deformed computer 
codes. Since technical details of such calculations are presented in Ref.\ 
\cite{OATDPDF.25} we focus here on the features which are relevant for the 
present study.

     The most of the calculations are carried out with the DD-MEZ, NL5(Z) and 
PC-Z CEDFs representing the DDME, NLME and PC classes of the functionals.
These functionals were developed in Ref.\ \cite{OATDPDF.25}  using global  
optimization. Separable pairing interaction of Ref.\ \cite{TMR.09} with globally 
optimized strength of pairing (see Ref.\ \cite{TA.21} and Eqs. (2) and (3) in Ref.\ 
\cite{OATDPDF.25})  is used in the pairing channel.

  The employed computer codes have been substantially modified: they
were converted to the Fortran F95 standard which allows better memory 
management and some other changes were implemented. As a result, the 
calculations in extremely large fermonic and bosonic bases become possible. 
At  present, in both computer codes one can achieve the solution corresponding 
to infinite bosonic base (see Ref.\ \cite{OATDPDF.25}). The $N_B=40$ basis
is used in the present paper following the recommendation of Ref.\ 
\cite{OATDPDF.25}.

As illustrated below, one can achieve the numerical solution corresponding to infinite fermionic 
base in spherical RHB code which allows benchmarking of the solutions in 
smaller basis or spherical solutions in axially deformed RHB code.  The numerical solution 
corresponding  to infinite fermionic basis are also achievable in many light and 
medium mass  nuclei in axially deformed RHB code for meson exchange functionals
but still the extrapolation procedures to such bases discussed in Refs.\ 
\cite{TOAPT.24,OATDPDF.25} are required for the majority of  heavy nuclei. 
The later is due to the fact that axially deformed RHB calculations with separable
pairing employed in the present study can be carried out only up to $N_F=40$ on available computers
(see Ref.\ \cite{OATDPDF.25}). However, the use of simpler pairing (for 
example, the monopole one) or switching off the pairing altogether allows
the calculations in such a code  for fermionic bases extended up to $N_F=60$
because of the memory reduction as compared with the case of separable
pairing (see discussion of Fig.\ 2 in Ref.\ \cite{OATDPDF.25}). This represents 
an alternative way for the test of the convergence of binding energies as a 
function of $N_F$ in the nuclei with deformation. 

\begin{figure}[htb]
 \begin{center}
   \centering
   \includegraphics*[width=8.5cm]{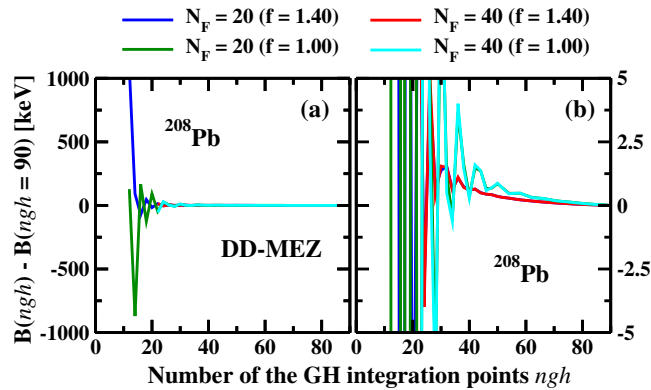}
   \caption{The dependence of binding energies on the number of the GH integration 
points $ngh$ in the ground state of $^{208}$Pb. The 
calculations are performed for indicated combinations of $N_F$ and scaling factor 
$f$. The right panel shows the results presented in the left one but in significantly 
reduced energy window. The $ngh=90$ solution corresponds to the exact one. 
\label{spher-ngh-208Pb} 
}
\end{center}
\end{figure}

   The computer codes (spherical, axially deformed \cite{AARR.14}, triaxial 
deformed  \cite{AAR.10}, triaxial cranking \cite{AKR.96,CRHB} and axial octupole 
deformed \cite{AAR.16})  employed and developed  by our  group  show the same 
convergence of binding energies as a function of $N_F$.  In Ref.\ \cite{TOAPT.24} 
we verified  for selected set of spherical and  deformed nuclei that the codes developed 
in our group and those existing in the DIRHB package of the RHB codes (see Ref.\ 
\cite{DIRHB-code.14}) provide almost the same (within a few keVs) results for binding 
energies as  a function of $N_F$ and $N_B$. 

\section{The dependence of the results on the number of integration 
points in Gauss-Hermite and Gauss-Laguerre quadratures}
\label{Quadratures}                                  

    The Gauss-Hermite (GH) and Gauss-Laguerre (GL) quadratures
\cite{NumRec}  are most  frequently used in numerical integration in the CDFT 
framework (see Refs.\  \cite{GRT.90,DIRHB-code.14}). Despite that no detailed 
analysis of these procedures and their numerical accuracy in the context of the
CDFT applications has been published so far. To fill this gap in our knowledge let 
us start from spherical nuclei. The integration in that case is defined 
by the number of the GH integration points (labelled as {\it ngh} in computer codes). 
It  is well recognized in the CDFT community that the accuracy of the GH integration is 
dependent on  {\it ngh}, but the fact that it also defines the size of the nucleus and 
the number of the GH integration points per unit of length is overlooked. 
 
     Fig.\ \ref{208Pb-density} illustrates the latter features by comparing the 
densities at the location of the GH integration points obtained in the calculations with 
$ngh=12$, 30 and 90. Note that due to numerical constraints the first value has 
frequently been used in the calculations at early stages of the CDFT development in 
the 80s and 90s  of the last century. One can see that the last GH integration point, 
which defines approximate size of the nucleus in the calculations,  is located at 
$r\approx 12.5$,  21.0 
and  35.4 fm in the calculations with $ngh=12,$ 30 and 90, respectively.  Thus, the 
use of higher $ngh$ value in the calculations leads to a better accounting of the low 
density tail of the density distribution.

    In addition, one can see that the number of the GH integration points per unit of 
length drastically increases on transition from $ngh=12$ to $ngh=90$. The 
single-particle wave functions show oscillatory behavior as a function of radial 
coordinate which depends on their nodal structure (see, for example,  Fig.\ 2 in 
Ref.\ \cite{PA.23}) and this behavior is better accounted in the integration for 
a larger number of the GH integration points per unit of length.

\begin{figure}[htb]
 \begin{center}
   \centering
      \includegraphics*[width=8.5cm]{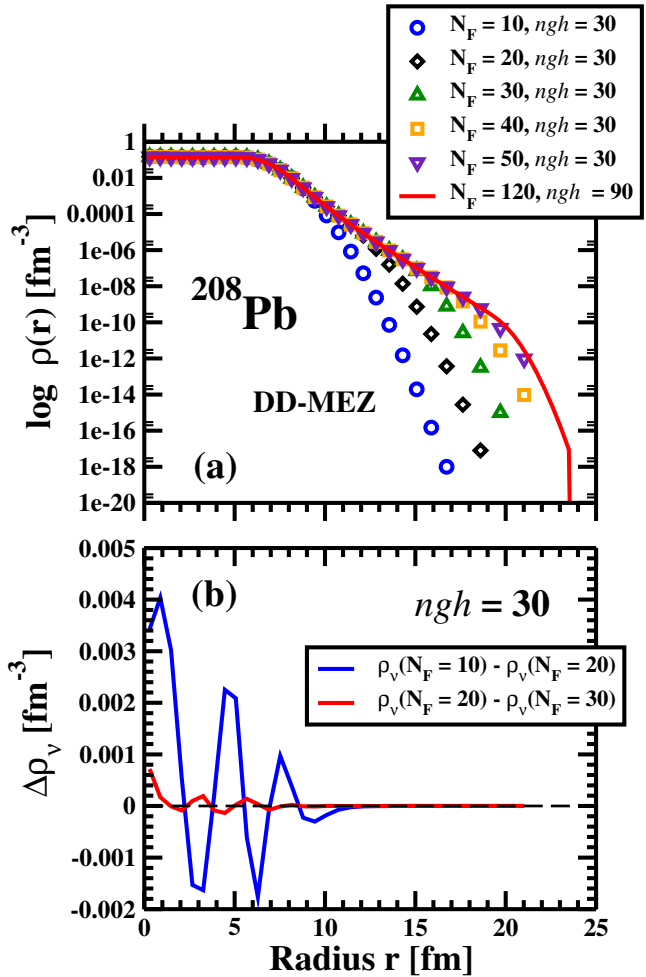}
   \caption{(a) Neutron densities at the GH integration points obtained in 
the calculations with $ngh=30$  and $N_F=10$, 20,  30,  40 
and 50 compared 
with exact solution  shown by red line. (b) The differences $\Delta \rho_{\nu}$
of neutron densities at the GH integration points obtained in the calculations with
$ngh=30$ and indicated values of $N_F$.  The calculations
are carried out with scaling factor $f=1.4$.
\label{spher-dens-208Pb} 
}
\end{center}
\end{figure}

The combined effect of these two factors is clearly visible in Fig.\ \ref{spher-ngh-208Pb}.
The calculated binding energy 
deviates from exact solution by no more than 2 keV above some critical value 
$ngh_{crit}\approx 40$. Moreover, above this value it gradually approaches exact one 
with increasing $ngh$.  This is due to two factors. First, the density of the GH integration 
points per unit of length which raise with increasing $ngh$ becomes sufficiently large in 
the interior and surface region of the nucleus so that its further increase does not improve 
the accuracy of the description of the wave function. Second, in the tail of the density 
distribution of the nucleus many of the GH weights are so small that corresponding terms 
of the GH quadrature contribute negligibly to the result (see Ref.\ \cite{Tref.21}). 

   An oscillatory behavior of the $B(ngh)-B(ngh=90)$ is seen around $ngh\approx 40$ and 
it increases in magnitude with decreasing $ngh$ (see Fig.\ \ref{spher-ngh-208Pb}):
the difference between $B(ngh)$ and exact solutions reaches the vicinity of 1 MeV at low
values of $ngh$. These features are due to oscillatory behaviour of the GH and GL 
quadrature errors as a function of $ngh$ (see Refs.\ \cite{Tref.21,Davies-book}). 
The magnitude of these oscillations depends on the scaling factor $f$. For example,
such oscillations are more pronounced in the $f=1.00$ results as compared with the $f=1.40$ 
ones (see Fig.\ \ref{spher-ngh-208Pb}(b)). Moreover, these two results oscillate opposite to 
each other.  In contrast, the convergence curve almost does not depend on $N_F$
for a given $f$ value. Note that the calculations for $N_F=40$ are numerically unstable for 
$ngh\leq 22$ and thus are not shown in Fig.\ \ref{spher-ngh-208Pb}. 

   Detailed investigation of the convergence of the binding energies as a function of 
 $ngh$ has also been performed in spherical $^{48}$Ca, $^{132}$Sn and $^{304}$120 nuclei 
as well as in normal-deformed  (with quadrupole deformation $\beta_2 \approx 0.3$) $^{240}$Pu nucleus. In the latter case, the investigation of the 
convergence has been carried out as a function of the number of the GH ($ngh$) and 
GL ($ngl$) integration points with respect of an exact solution with $ngh=90, ngl=90$.
The convergence pattern of Fig.\ \ref{spher-ngh-208Pb} is seen in all these cases: 
spherical results with $ngh=40$ and deformed ones with $ngh=ngl=40$ 
reproduce exact results with accuracy better than 2 keV.  These values of $ngh$ and $ngl$ 
are used in all studies presented below. If such high an accuracy is not required then 
the $ngh=30$ and $ngh=ngl=30$ sets provide acceptable accuracy in the spherical
and axially deformed RHB calculations. For example, such values were used  in the studies 
of Refs.\ \cite{TOAPT.24,OATDPDF.25}). 

  Above discussed errors in the calculations of binding energies is a reason 
why in standard GH and GL integrations one should go substantially outside of the nucleus 
to achieve numerically accurate results. 
 However, then the part of the integration space 
corresponding to extremely low densities emerges and it is expected to contribute only 
marginally to final results. This deficiency of the standard Gauss quadratures is known 
(see Ref.\ \cite{Tref.21,Davies-book}) but, to our knowledge, has not  been 
discussed  in nuclear DFT. While it is not very critical for spherical or axially
deformed nuclei, it becomes more important in the nuclei (such as triaxial ones)
the calculation of which requires 3-dimensional integration. Adaptive
Guass quadratures \cite{JA.20,PPAB.20} which dynamically adjusts the nodes and weights to specific 
features of the nucleus could potentially eliminate this deficiency (i.e. substantially
decrease the integration volume)
and thus considerably reduce computational time.

  The value of $ngh$ to be used in the calculations is also dependent on the following 
considerations. Finite HO basis in nuclear many-body  calculations effectively imposes a 
hard-wall boundary conditions in coordinate space,  i.e. it is equivalent to a spherical 
cavity of a radius $L_0$ \cite{FHP.12,BEHPW.16} 
\begin{eqnarray} 
L_0 = \sqrt{2(N_F+3/2)b}.  
\label{radius}
\end{eqnarray} 
in the case of spherical nuclei. The radius of this cavity defined by $\hbar \omega_0$ and $N_F$ 
of the employed HO basis should be larger than the radius $r$ of the nucleus. Here, 
$b=\sqrt{\hbar/(m\omega_0)}$  is the oscillator length of the basis and $m$ denotes the 
nucleon mass. Eq.\ (\ref{radius})  provides a rough estimate (see Ref.\ \cite{MEFHP.13}) and 
in practical CDFT calculations the two-dimensional $(ngh, N_F)$ space should be explored 
to define the boundaries beyond which the increase of either of these parameters does not 
change numerical solution.

    Indeed,  Fig.\ \ref{spher-dens-208Pb}(a) clearly shows that the correct reproduction 
of low density tail of the neutron density distribution at large radial coordinate requires sufficiently 
large fermionic basis. The calculations with $N_F=10$ substantially underestimate the exact
solution for $r\geq 10$ fm.
The increase of the basis to $N_F=30$ reproduces the low-density 
tail up  to $r\approx 15$  fm but underestimates exact solution at higher radial coordinates. The 
solution with  $N_F=50$  comes very close to an exact one.

\section{The convergence of the nuclear binding energies as a function of size of fermionic
             basis: the examples of spherical $^{208}$Pb and deformed $^{240}$Pu nuclei}
\label{conv-samples}

\subsection{Meson exchange functionals}

\begin{figure}[htb]
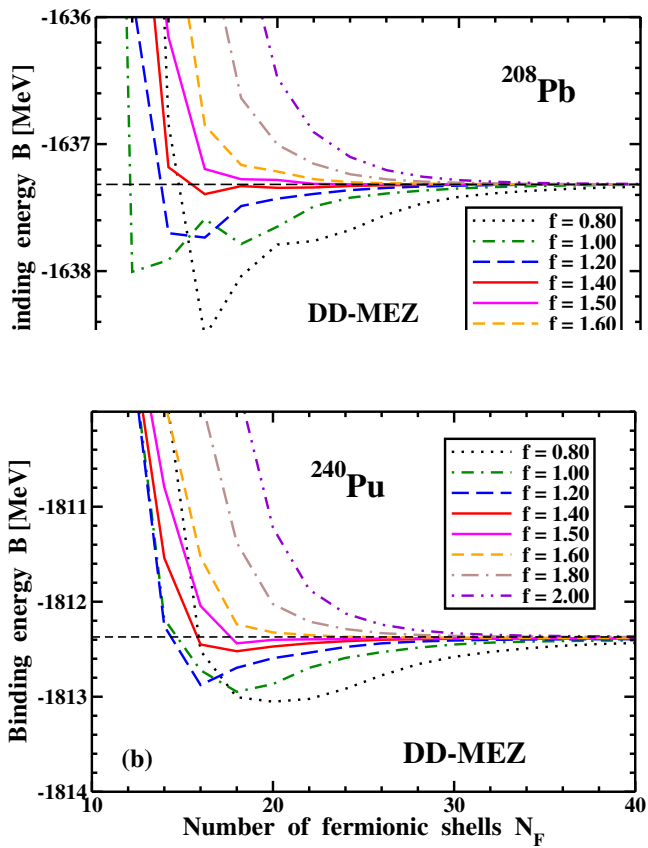

	\centering
\includegraphics[width=8.4cm]{fig-4-a.eps}
\includegraphics[width=8.5cm]{fig-4-b.eps}
\caption{The binding energies of the ground states of the $^{208}$Pb and 
  $^{240}$Pu nuclei as a function of $N_F$ for different values of scaling
  factor $f$. Thin dashed 
line shows the exact value of binding energy corresponding to infinite basis.
\label{conver-208Pb-240Pu}
}
\end{figure}

Fig.\ \ref{conver-208Pb-240Pu} shows the dependence  of binding energies of 
the ground states of spherical  $^{208}$Pb and normal-deformed $^{240}$Pu nuclei as a 
function of  $N_F$ for different values of scaling factor  $f$ ranging from 0.8 up to 2.0 for 
the DD-MEZ functional. These two nuclei share similar features discussed below.

\begin{figure}[htb]
	\centering
\includegraphics[width=8.5cm]{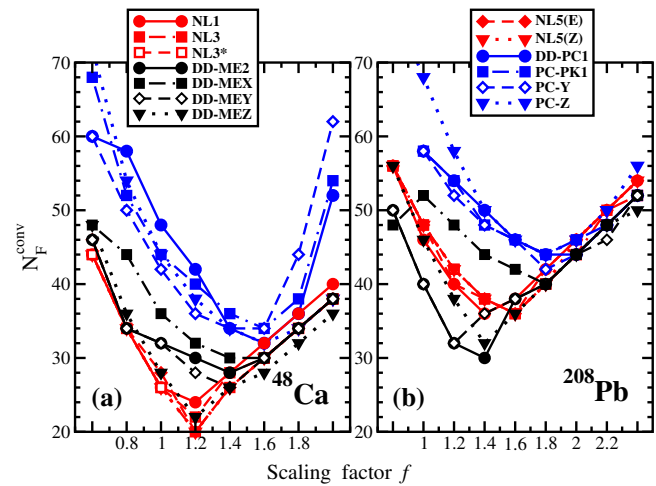}
\caption{The values of $N_F^{conv}$ at which the calculations converge as a function of 
scaling factor $f$ for indicated functionals. The convergence point $N_F^{conv}$ is reached 
when $|B(N_F^{conv}) - B(N_F=90)| \leq \varepsilon$ where $\varepsilon$ is numerical
accuracy of the calculations of binding energy in variational calculations ($\varepsilon$ =1 keV
in our case). The NL1, NL3, NL3*, NL5(E) and NL5(Z) CEDFs belong to the NLME class
of the functionals.  The DDME class of the functionals is represented by the DD-ME2, DD-MEX,
DD-MEY and DD-MEZ CEDFs. The DD-PC1, PC-PK1, PC-Y and PC-Z ones
are representatives of the PC class of the CEDFs.
\label{conver-NF}
}
\end{figure}

   Let us first consider the $^{208}$Pb results. The calculations with $f=0.8$, 1.0, 
1.2, 1.4, 1.5, 1.6, 1.8 and 2.0 fully converge  to the same binding 
energy at $N_F^{conv}=52$, 46, 38, 32, 36, 42, 
and 48, respectively (see Fig.\ \ref{conver-NF}). 
The studies of spherical $^{48}$Ca and $^{304}$120 nuclei reveal 
the same features of the convergence as those in $^{208}$Pb. 
 
  The calculation with $N_F>40$ are impossible in axial RHB code with separable pairing.
It is only in the 
calculations with $f=1.4$  and 1.5 that full convergence of binding energies in $^{240}$Pu is 
reached at $N_F^{conv}=34$. The values higher than $N_F=40$ are required for
full convergence in the calculations with other $f$ values. 
 However, the binding energy curves 
are monotonic above $N_F\approx 22$ in $^{240}$Pu for all $f$ values
(see Fig.\ \ref{conver-208Pb-240Pu}(b)).
Thus, using extrapolation procedure outlined in Sec.\ VI of Ref.\ \cite{TOAPT.24} the 
binding energies  corresponding to infinite fermionic basis have been obtained for the 
$f$ values for which full convergence has not been reached at $N_F=40$. The convergent
and extrapolated solutions differ by only few keVs.

   The convergence of binding energies as a function of $N_F$ depends 
on scaling factor $f$ (see Fig.\ \ref{conver-208Pb-240Pu}). For scaling 
factors $f=1.6$, 1.8 and 2.0 the convergence curve is monotonic i.e. the nucleus 
always 
becomes more bound with increasing $N_F$. We label this feature as {\it pattern A}
convergence. For other values of $f$, the convergence curve is
non-monotonic i.e. the 
binding increases rapidly with increasing $N_F$ at low $N_F$, then the nucleus becomes 
more bound than the exact solution in transitional region, and only  with  further increase 
of $N_F$ it monotonically approaches the exact solution from below by getting less 
bound.  This feature is labelled as {\it pattern B} in further discussion.
It is consistent with the fact that in some physical systems the convergence curve starts to 
behave asymptotically (i.e. monotonically) only above some critical size of the basis (see 
introduction in Ref.\ \cite{CAKKMV.12} and example quoted as reference [16] in this paper).

   It is frequently stated in the literature that the nucleus gets more bound with the 
increase of the size of the basis in the calculations. Indeed, this is seen in a number
of publications (see, for example, Refs.\ \cite{MVS.09,FHP.12}). However, in general 
the variational principle guarantees only an extremum which could be a stationary 
point rather than a minimum (see note appearing as Ref. [32] in Ref.\ \cite{CAKKMV.12} 
and examples discussed in this note).  Such examples are seen for effective 
interactions  in no-core shell model calculations: the convergence to exact solution 
could be from above, from below or oscillatory (see discussion of Fig.\ 1 in Ref.\ 
\cite{NQSB.09}). Note that {\it pattern B} convergence is also seen in {\it ab-initio} 
calculations of light nuclei (see, for example, Figs. 4 and 8 in Ref.\ \cite{CAKKMV.12}).
 
   The situation becomes even more complicated  in the CDFT which contains 
two bases (fermionic and bosonic) with the convergence affected by the coupling 
between  them (see Sec.\ \ref{Interconnection} below).  Thus, dependent on scaling factor $f$ of the 
oscillator frequency one can observe both patterns (A and B) of the convergence 
in the same nucleus (see Fig.\ \ref{conver-208Pb-240Pu}).  
This also indicates  that the results obtained in the non-relativistic framework 
which has only one  basis (fermionic) should not be extrapolated to relativistic 
one without verification. 

   The difference in the rate of the convergence (i.e. the slope of binding energies as a 
function of $N_F$) in the $N_F=10-20$ and $N_F=20-30$ regions seen in Fig.\ \ref{conver-208Pb-240Pu} 
 is easy to understand from Fig.\ \ref{spher-dens-208Pb}(b). It is 
unreasonable to expect that the $N_F=10$ basis provides an accurate description of 
$^{208}$Pb. The increase of the size of the basis provides a richer and more complete
mathematical space to describe nuclear wave function. This leads to appreciable
change of neutron density on the transition  from $N_F=10$ to $N_F=20$ [i.e. 
$\rho_{\nu}(N_F=10) - \rho_{\nu}(N_F=20)$, see \ref{spher-dens-208Pb}(b)] which 
explains a large slope of binding energy as a function of $N_F$ in the $N_F=10-20$
range (see Fig.\ \ref{conver-208Pb-240Pu}).
Further increase of the basis by 10 fermionic shells triggers substantially smaller changes 
of neutron densities (see $\rho_{\nu}(N_F=20) - \rho_{\nu}(N_F=30)$ curve in Fig.\
\ref{spher-dens-208Pb}(b)) and this explains why binding energy changes in the $N_F=20-30$ 
region are rather small. The results presented on
Fig.\ \ref{spher-dens-208Pb}(b) strongly suggest that the changes of the convergence rate 
with the increase of $N_F$ are predominantly driven by the density changes in the interior  of the 
nucleus and not by the build up of the low density tail at large radial coordinate seen 
in  Fig.\ \ref{spher-dens-208Pb}(a). 
  
   Systematic numerical analysis of the CDFT results shows that the binding 
energies converge in a monotonic
way in the pattern A.  However, the same is true for the pattern B convergence but only
for the $N_F$ values above some critical value $N_F^{crit}$. For example, one can see in
Fig.\ \ref{conver-208Pb-240Pu}(a) that the binding energy curves for $f=0.8$ and $f=1.0$
converge monotonically
above $N_F^{crit}=16$ and $N_F^{crit}=18$, respectively.

    The investigated cases of the $^{48}$Ca, $^{208}$Pb, $^{240}$Pu 
and $^{304}120$ nuclei clearly indicate that there is an optimal value of scaling factor 
$f$ which (i) provides the fastest convergence to an exact solution and  (ii) for which relatively 
small basis (as compared with other $f$ values) is needed to obtain highly accurate 
reproduction of an exact solution.  For the DD-MEZ functional,  $f=1.5$  represents 
such  an optimal value for heavy nuclei for $N_F\geq 16$ with only slightly worse accuracy 
provided by $f=1.4$. However, comparable good accuracy is obtained  at  $f=1.3$
for $N\geq 12$ in the $^{48}$Ca nucleus. This is illustrated in Fig.\ \ref{conver-208Pb-240Pu} 
which shows that the $f=1.5$  solution is the closest to the exact one among 
considered solutions in $^{208}$Pb and $^{240}$Pu. Moreover, it is consistently close to the 
exact solution starting from $N_F=16$: larger basis is needed to achieve  the same accuracy 
of the reproduction of exact solution for other $f$ values.

  One can also ask a question on how above discussed features depend on the
functional. It turns out that they are generic for a given class of the
functionals. The analysis of the convergence properties of the $^{48}$Ca and 
$^{208}$Pb nuclei  carried out with DD-ME2 \cite{DD-ME2}, DD-MEX \cite{TAAR.20} 
and DD-MEY \cite{TA.23} functionals shows the same features as those seen
for the DD-MEZ CEDF (see Fig.\ \ref{conver-NF}).

    It turns out that  similar features exist also for the NLME functionals. This conclusion 
 is born out in the calculations carried out with NL1 \cite{NL1}, NL3 \cite{NL3}, NL3* \cite{NL3*}, 
 NL5(E) \cite{AAT.19} and NL5(Z) \cite{OATDPDF.25} functionals  for the $^{48}$Ca and 
 $^{208}$Pb nuclei (see Fig.\ \ref{conver-NF}). However, the best agreement with exact  
 binding energies is obtained for 
 $f=1.2$ both in $^{48}$Ca (for $N_F\geq 12$)
  and in $^{208}$Pb (typically for $N\geq 18$).

\subsection{Point coupling functionals}
\label{PC-functionals}

\begin{figure}[htb]
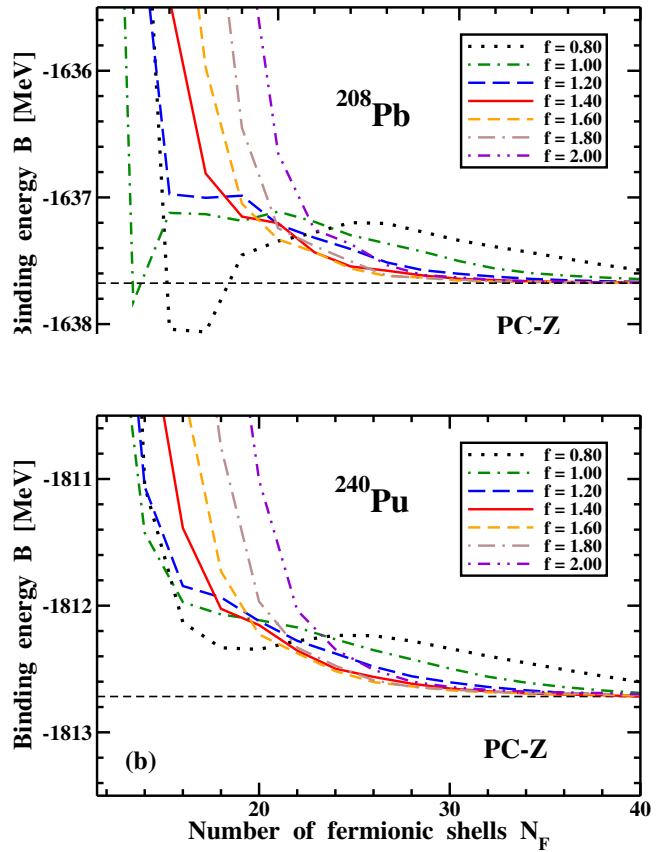

	\centering
\includegraphics[width=8.5cm]{fig-6-a.eps}
\includegraphics[width=8.5cm]{fig-6-b.eps}
\caption{The same as Fig.\ \ref{conver-208Pb-240Pu} but for the PC-Z functional.
Thin dashed line shows the exact value of binding energy corresponding to infinite 
basis in the case of $^{208}$Pb and the binding energy of the $f=1.4$ solution at  
$N_F=40$ in the case of $^{240}$Pu. The latter is the lowest one among considered
solutions at $N_F=40$.
\label{conver-208Pb-240Pu-pcz}
}
\end{figure}

     The convergence of calculated binding energies of the $^{208}$Pb and 
 $^{240}$Pu nuclei for the PC-Z functional is shown in Fig.\ \ref{conver-208Pb-240Pu-pcz}. 
In $^{208}$Pb, one can see pattern A convergence for the $f=2.0$, 1.8, 1.6 and 1.4 values. 
The $f=1.2$ curve comes very close to this pattern. The $f=1.0$ and 0.8 curves experience 
some  disturbances at $N_F=12$ and at $N_F=14$ and 16, respectively.  In deformed 
$^{240}$Pu nucleus, the pattern A convergence is obtained for all $f$ values with exception
of $f=0.8$ (see Fig.\ \ref{conver-208Pb-240Pu-pcz})(b).

   Fig.\ \ref{conver-208Pb-240Pu-pcz} shows that for $N_F\geq 18$, the convergence 
to exact solution proceeds from above with increasing $N_F$ for all values of $f$. 
Such features are also seen in the calculations with the DD-PC1, PC-PK1 and PC-Y
functionals for the $^{48}$Ca and $^{208}$Pb nuclei\footnote{The calculations
carried out with $f=1.0$ for spherical $^{40}$Ca, $^{132}$Sn and $^{304}$120 and 
deformed $^{42}$S, $^{164}$Dy and $^{270}$Ds nuclei with the DD-PC1 and 
PC-PK1 functionals and deformed $^{240}$Pu nucleus with the PC-Z one
also show the same features (see Figs. 5 and 6 in Ref.\ \cite{TOAPT.24} and Fig. 2 
in Ref.\ \cite{OATDPDF.25}).}  so they 
are generic for a given class of the functionals. This is very  similar to many non-relativistic 
calculations (see, for example, Refs.\ \cite{MVS.09,FHP.12}). However, this is in contrast to 
covariant meson exchange (ME) functionals discussed in previous subsection which converge 
to an exact  solution either from below of from above dependent on scaling factor $f$.

   These features have important consequences. In contrast to the ME functionals
there is no optimal value of $f$ which provides almost perfect reproduction of exact results
at moderate $N_F\approx 20$ in the PC functionals. For example, dependent on scaling factor $f$
the difference between  truncated (at $N_F=20$) and exact solutions is around 400 keV or
 more for the PC functionals  (see Fig.\ \ref{conver-208Pb-240Pu-pcz}). 
In contrast, by fine tuning of $f$ one can reduce this difference to almost zero in the ME functionals 
(see Fig.\ \ref{conver-208Pb-240Pu} and its discussion).  As a consequence, full 
convergence is typically reached at substantially higher values of $N_F$ in the PC functionals as 
compared with the ME ones (see Fig.\ \ref{conver-NF}).

One can also ask a question what would be the recommended value of $f$ which provides the 
smallest difference between exact and truncated results at moderate values of $N_F$. The 
analysis of the convergence curves shown in Fig.\ \ref{conver-208Pb-240Pu-pcz} indicates 
that the $f=1.4$, 1.6 and 1.8 values provide the fastest and comparable convergence to 
exact results at $N_F  \geq 20$. The same conclusion is obtained in the calculations with the
PC-PK1, PC-Y and DD-PC1 functionals. However, the analysis of the convergence in 
$^{48}$Ca carried out with four employed PC functionals indicates that the values of $f=1.2$, 1.4 and 
1.6 provide the fastest convergence. Moving away from these values slows down the
convergence and substantially increases the values of $N_F^{conv}$ at which full convergence
is reached (see Fig.\ \ref{conver-NF}).

\section{The oscillator frequency $\hbar \omega_0$ of the basis as a variational 
             parameter}
\label{variat-omega}

\begin{figure}[htb]
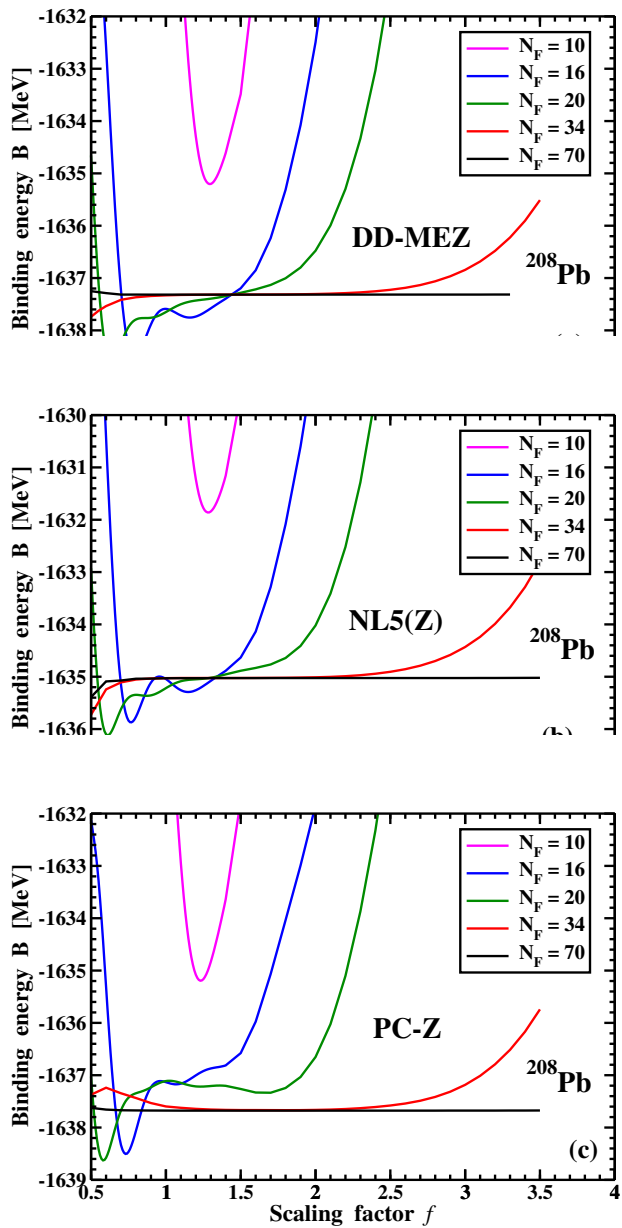

	\centering
\includegraphics[width=8.1cm]{fig-7-a.eps}
\includegraphics[width=8.1cm]{fig-7-b.eps}
\includegraphics[width=8.1cm]{fig-7-c.eps}
\caption{Binding energy of the ground state of the $^{208}$Pb nucleus as a function 
of $f$ for indicated functionals.  The exact  solution is represented by the $N_F=70$ results.
\label{dep-on-f}
}
\end{figure}

     A short review on convergence properties of the HO basis set expansions
in different theoretical frameworks presented in the introduction of Ref.\ \cite{CAKKMV.12} 
shows that there are two types of the approaches to the issue of the use of the 
parameters of the basis as variational ones.  In the first approach the oscillator 
frequency $\hbar \omega_0$ of the basis is used as a variational parameter but 
in another one it is fixed.  For example, the first approach has been used in earlier 
Skyrme DFT  calculations  in small basis  (see, for example, Refs.\ \cite{V.73,FQV.73}).  
The no-core shell model analysis shows that such calculations benefit from the treatment 
of $\hbar \omega_0$ as a variational parameter (see Fig. 6 in Ref.\ \cite{FHP.12}). Other 
examples can be found in Ref.\ \cite{CAKKMV.12} and references quoted therein.
 
   The CDFT represents an example of another approach: to our knowledge 
the oscillator frequency $\hbar \omega_0$   has never been used as a variational 
parameter in numerical calculations in its framework. Thus, it is important to understand whether the
treatment of $\hbar \omega_0$ as a variational parameter can be useful. To address 
this question we plot calculated nuclear binding energy of the ground state of 
$^{208}$Pb as a function of scaling factor $f$ of the oscillator frequency in Fig.\ 
\ref{dep-on-f} for the DDME,  NLME and PC classes of CEDFs.  The exact 
solutions are obtained for $N_F=70$: their numerical values are independent 
of scaling factor $f$ for a large range of $f$.  The optimization  of the $N_F=10$ 
solution with respect of $f$ brings calculated binding energy closer to the exact 
solution. However, there is no benefit in treatment of oscillator frequency as a 
variational parameter in the $N_F=16$ and $N_F=20$ calculations since the 
minimum of binding energy curves at $f\approx 0.7$ deviates more from exact 
solutions than the binding energies calculated at $f \approx 1.4-1.6$. One can also 
see that for all functionals the $N=34$ solution comes extremely closely to the 
exact one for $f\approx 1.2-2.0$.

  The analysis of Figs.\  \ref{conver-208Pb-240Pu} and \ref{dep-on-f}
suggests that the use of $\hbar \omega_0$ as a variational parameter is justified 
only for relatively low values of  $N_F$  which are no longer used in the state-of-the-art 
calculations.  For large $N_F$ values such an approach leads to disadvantages. 
Using the results presented in Figs.\ \ref{conver-208Pb-240Pu} as 
an illustration,  one can see that the optimization of the binding energies with respect of $\hbar \omega_0$ 
will lead to the envelope (as a function of $N_F$) of the lowest in energy solutions 
shown in this figure. However, this envelope deviates from exact solutions 
even at high $N_F$. This is because the solutions with $f\approx 0.7$ are substantially 
lower than the exact solutions for all functionals under study (see Fig.\ \ref{dep-on-f}). 

    These observations are in line with the analysis presented in the introduction of 
Ref.\ \cite{CAKKMV.12}  which indicates that only for small bases it is beneficial 
to use $\hbar\omega_0$ as variational parameter while for large ones such 
an approach does not provide any benefits.

     The analysis of Figs.\  \ref{conver-208Pb-240Pu} and \ref{dep-on-f}  
suggests alternative approach in which the oscillator frequency of the basis is
selected at moderate $N_F'\geq N_F^{crit}$ in such a way that the $N_F'$ solution reproduces
well exact solution at $N_F=\infty$. Since for a given $f$ the binding energies 
above  critical value $N_F^{crit}$ behave monotonically as a function of $N_F$ 
this guarantees that  (i) the difference $|B(N_F')-B(N_F=\infty)|$ provides the upper 
limit for the discrepancy between theory and experiment and (ii) that this difference 
reduces with the increase of $N_F$ above $N_F'$.  The basic idea behind this approach 
is illustrated by the $f=1.4$ binding energy curves in Fig.\  \ref{conver-208Pb-240Pu}: 
for this value of scaling factor $|B(N_F'=16)-B(N_F=\infty)|\approx 50$ 
keV both in $^{208}$Pb and $^{240}$Pu. This energy difference is substantially smaller 
as compared with that obtained when $\hbar \omega_0$ is treated as variational 
parameter.

\section{Impact of the coupling of fermionic and bosonic bases on
              convergence of binding  energies}
\label{Interconnection}                                  

\begin{figure}[htb]
 \begin{center}
   \centering
   \includegraphics*[width=8.5cm]{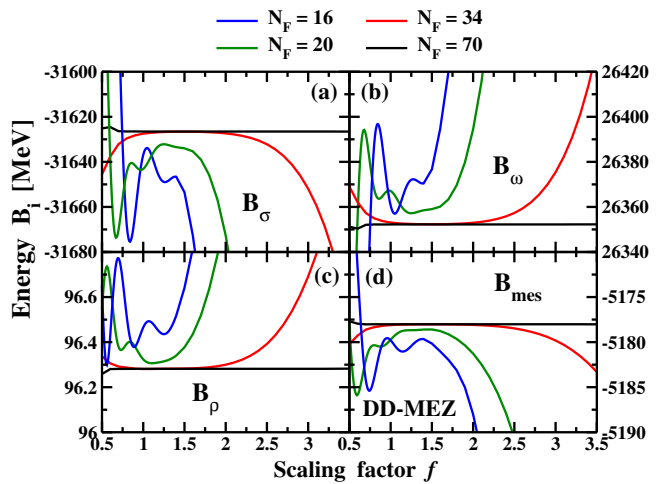}
   \caption{The dependence of mesonic energies $B_i$ on the scaling factor $f$
   in $^{208}$Pb.
\label{bos-fer-fine} 
}
\end{center}
\end{figure}

   The results presented in Sec.\ \ref{conv-samples} clearly illustrate principal
difference between the PC and ME functionals. In both classes of the 
functionals, the convergence to exact solution is monotonic above some 
critical $N_F^{crit}$ value (for example, above $N_F^{crit}\approx 20$ in $^{208}$Pb
and $^{240}$Pu [see Figs.\ \ref{conver-208Pb-240Pu} and \ref{conver-208Pb-240Pu-pcz}]). 
However, similar to many
non-relativistic functionals the convergence to exact solution for $N_F\geq N_F^{crit}$
is from above for all employed $f$ values in the PC CEDFs.  In contrast,  the convergence 
to exact solution can be either from above or from below dependent on scaling factor $f$ 
in the ME functionals.
   
   The principal difference of these two classes of CEDFs lies in their structure.
There are no mesons in the PC functionals. As a result, there is only one (fermionic) 
basis and the convergence of binding energies  behaves similarly to non-relativistic 
theories. In contrast,  the ME functionals contain fermions (nucleons) and 
bosons (mesons) which leads to a unique two bases (fermionic and bosonic)
structure of quantum system.  It is reasonable to expect that if these two bases 
would be completely decoupled then the convergence in the fermionic basis would 
be similar to that of the PC functionals. However, as discussed below this is not 
a case and  the dependence of the convergence of binding energies on the scaling 
factor $f$ differs substantially in the ME functionals as compared with the PC ones.
This is a unique feature of the ME functionals which allows to  reproduce very 
accurately the exact solution with relatively moderate fermionic  basis by selecting
fixed optimal scaling factor $f$. In contrast, to achieve comparable accuracy substantially 
larger basis in required for the PC functionals.
   
The mesons are present in the ME functionals but absent in the PC ones and
this is a reason for above discussed differences in the convergence.     
The binding energy $B_{mes}$ of  even-even nuclei in the mesonic 
sector of the CDFT in the laboratory frame is  given by
\begin{eqnarray}
B_{mes} = B_{\sigma} + B_{\omega} + B_{\rho} + B_{{\sigma}NL} 
\label{binding-bos}
\end{eqnarray}
for the ME functionals. Here $B_{\sigma}$ and $B_{\omega}$ are attractive and repulsive 
energies due to the $\sigma$ and $\omega$ mesons, respectively.  $B_{\rho}$ and 
$B_{\sigma NL}$ are the energies due to the $\rho$-meson\footnote{For the 
isovector–vector $\rho$-meson the time-like components give rise to 
a short range  repulsion for like particles (pp and nn) and a short range attraction for unlike 
particles (np).} and nonlinear contribution to the energy of the isoscalar-scalar $\sigma$-field 
\cite{BB.77}, respectively. Note that the latter term is present only in the NLME functionals.

\begin{figure}[htb]
\begin{center}
\centering
\includegraphics*[width=8.5cm]{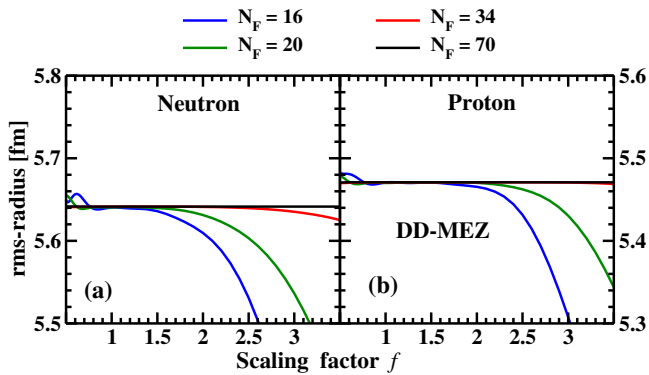}
\caption{Neutron and proton rms radii as a function of scaling factor $f$ for
                 indicated values of $N_F$.
\label{bos-fer-radii} 
}
\end{center}
\end{figure}

    They are defined as (see Refs.\ \cite{CRHB,AA.10} for guidance)
\begin{eqnarray}
B_{\sigma} & = & -\frac{1}{2} g_{\sigma} \int \sigma(\mathbf{r}) \rho_s(\mathbf{r})  d{\mathbf r}, \\
B_{\omega} & = &  -\frac{1}{2} g_{\omega} \int \omega_0(\mathbf{r}) \rho_v^{is} (\mathbf{r})  d{\mathbf r}, \\
B_{\rho} & = & -\frac{1}{2} g_{\rho} \int \rho_0(\mathbf{r}) \rho_v^{iv} (\mathbf{r})  d{\mathbf r}, \\
B_{{\sigma}NL} & = & -\frac{1}{2} \int \left[\frac{1}{3} g_2 \sigma^3(\mathbf{r}) + 
\frac{1}{2} g_3 \sigma^4(\mathbf{r}) \right]. 
\label{binding-bos-diff}
\end{eqnarray}
Here $\rho_s$, $\rho_v^{is}$, $\rho_v^{iv}$ and $r_v^p$ are fermionic scalar, fermionic 
isoscalar vector, fermionic isovector vector, and fermionic proton densities, respectively
(see Refs.\ \cite{CRHB,AA.10} for respective definitions).

\begin{figure}[htb]
 \begin{center}
   \centering
   \includegraphics*[width=8.5cm]{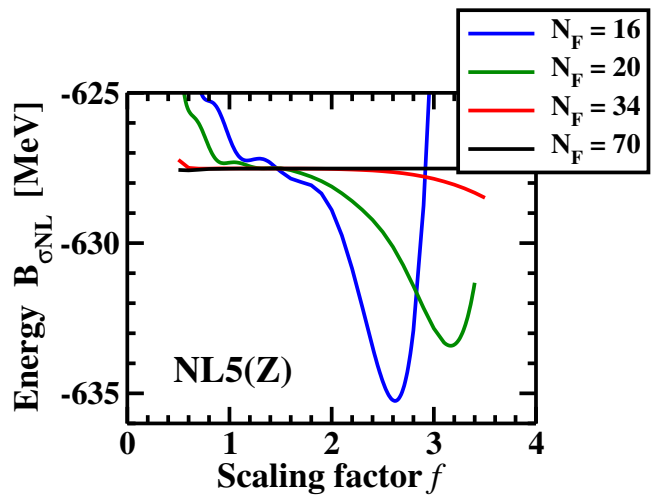}
   \caption{The dependence of the $B_{\sigma NL}$ energy on scaling factor 
   $f$ in $^{208}$Pb. 
\label{nonlinear-contr} 
}
\end{center}
\end{figure}

     Note  that mesonic fields $\sigma(\mathbf{r})$, $\omega_0(\mathbf{r})$ and
$\rho_0(\mathbf{r})$  are folded by these fermionic densities in the integrals 
defining $B^{\sigma}$,  $B^{\omega}$ and $B^{\rho}$. This strongly suggests 
that respective bosonic  energies should depend on the details of the calculations in the 
fermionic sector in the case of truncated fermionic basis. Indeed, the results presented in 
Fig.\ \ref{bos-fer-fine} confirm that. The source of these modifications is traced back to 
the changes of fermionic densities with increasing $N_F$ [see Fig.\ 
\ref{spher-dens-208Pb}(b)]: note that we use rms proton and neutron radii in Fig.\ 
\ref{bos-fer-radii} to illustrate these changes in  densities. By comparing the $N_F=20$ 
and $N_F=90$ results in Figs.\ \ref{bos-fer-fine}  and \ref{bos-fer-radii}  one can 
conclude that the differences between exact and truncated results for radii and 
bosonic energies are correlated. 

    Even very small modifications of fermionic densities reflect themselves in 
large changes of the $B_{\sigma}$ and $B_{\omega}$ energies and this is especially 
pronounced for the $\sigma$ meson [see Figs.\ \ref{bos-fer-fine}(a) and (b)]. This is 
because the $\sigma$ and $\omega$ mesons are responsible for the creation of attractive 
$S \approx -400$ MeV/nucleon and repulsive $V\approx +350$ MeV/nucleon potentials,
respectively (see Ref.\ \cite{Rei.89}). One can see that  with increasing $N_F$  the 
$B_{\sigma}$  and $B_{\omega}$ energies converge to the exact solution from below 
and above, respectively. Moreover, the truncation of the basis has a larger impact on 
$B_{\sigma}$ as compared with $B_{\omega}$. The impact of the truncation of fermionic 
basis on the $B_{\rho}$ is small. Thus, the convergence of total mesonic energy $B_{mes}$ 
is defined  almost entirely by the convergencies of $B_{\sigma}$ and $B_{\omega}$ and as 
one can see in Fig.\ \ref{bos-fer-fine}(d) it always proceed to the exact solution from below.

   Figs.\ \ref{bos-fer-fine}(d) shows that there is a pronounced dependence of
$B_{mes}$ on scaling factor $f$. For example, the $N_F=20$ solution comes closest
to the exact one for the $f$ values located between $\approx 1.15$ and $\approx 1.6$. 
The difference between these two solutions raises rapidly when the $f$ value moves
outside of this range. In contrast, the $N_F=34$ solution reproduces the exact one for 
a  broader range of the $f$ values but the deviation between these two solutions
still increase (but at slower rate as compared with $N_F=20$ case) outside of this range.

   Above discussed properties allow to understand unique features of the convergence of 
the ME functionals. The total mesonic energy $B_{mes}$ converges to exact solution 
from below. In contrast, fermionic energies converge from above similar to the PC 
functionals (see Fig.\ \ref{conver-208Pb-240Pu-pcz}). The convergence of both of these 
energies depend on scaling factor $f$ of oscillator frequency $\hbar \omega_0$.  As a result, 
by selecting  $f$ one can achieve that nuclear binding energy, which is a sum of fermionic 
and mesonic energies, converges to the exact solution either from below or from above 
or is nearly flat as a function of $N_F$ above some value $N_F^{crit}$ (see Fig.\ 
\ref{conver-208Pb-240Pu}).

   The  nonlinear contribution to the energy of the isoscalar-scalar $\sigma$-field  
$B_{\sigma NL}$  shows a different dependence on $N_F$ and scaling factor $f$
as compared with that seen in Fig.\ \ref{bos-fer-fine}. In the $f=0.8-2.0$ range, the 
convergence of $B_{\sigma NL}$ to the exact solution with increasing $N_F$
proceeds from above for $f\leq 1.5$ and from below for $f\geq 1.5$. This is a reason 
why the optimal values  of $f$ are typically lower for the NLME functionals as compared 
with the DDME ones (see earlier discussions and the results presented in Sec.\ 
\ref{global-convergence}).

\begin{turnpage}
\begin{figure*}[htb]
 \begin{center}
   \centering
   \includegraphics*[width=23.5cm]{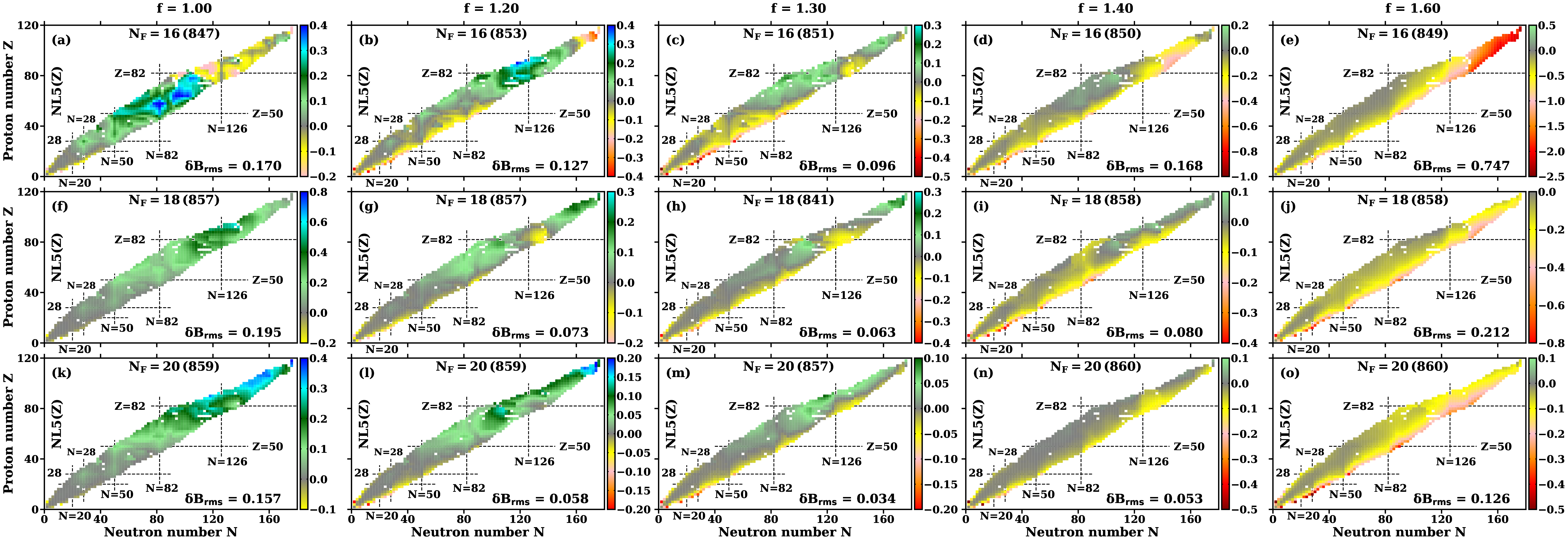}
\caption{The differences $B(N_F=\infty) -B(N_F)$ between binding energies of exact 
and truncated solutions for indicated values  of scaling factor $f$ and $N_F$ for the NL5(Z) 
CEDF. The global rms differences  $\delta B_{rms}$ (in MeV) between exact and truncated 
solutions are shown on each panel. Note that the ranges of colormaps are individual for 
each panel.  Only the nuclei for which calculated quadrupole deformations $\beta_2$ of 
exact and truncated solutions satisfy the condition $|\beta_2(N_F) - \beta_2(N_F=\infty)| \leq 0.05$
are used in the comparison: the number of such nuclei is indicated in parentheses after 
$N_F$ value on each panel.  This is done to avoid the comparison of the solutions in 
different closely lying minima (such as oblate and prolate) in which the increase of $N_F$
triggers the transition from one minimum to another or the solutions in the soft 
potential energy surfaces for which there is an appreciable  drift  of calculated 
$\beta_2$ values with increasing $N_F$. 
\label{global-NL5Z} 
}
\end{center}
\end{figure*}
\end{turnpage}

\section{Global analysis of the convergence errors for moderately sized fermionic
bases}
\label{global-convergence}
 
  Considered above cases of the optimization of the HO basis are 
restricted to a few nuclei. Thus, it is important to understand whether 
such an optimization works globally and how significant 
is an improvement over the results obtained with oscillator frequency 
$\hbar \omega_0 = 41 A^{-1/3}$ MeV. 

  To achieve that the exact\footnote{Such solutions are also labelled
as the $N_F=\infty$ ones in further discussion.} (within better than $\approx 10$ 
keV numerical error bar)\footnote{Note that the present 
global analysis is restricted to the DDME and NLME classes of
CEDFs since it is extremely numerically costly to get accurate solutions
corresponding to infinite fermionic basis in actinides and superheavy nuclei 
in axial RHB calculations for the PC functionals (see Fig.\ \ref{conver-NF} and 
Sec.\ \ref{PC-functionals} in the present paper and Refs.\ 
\cite{TOAPT.24,OATDPDF.25}).} solutions corresponding to infinite fermionic basis are 
compared with the ones calculated in the truncated $N_F=16$, 18 and 20 bases 
with scaling factors $f=0.8$, 0.9, 1.0, 1.1, 1.2, 1.3, 1.4, 1.5, 1.6, 1.7, and 1.8. To obtain the exact solutions 
the axial RHB calculations in the $N_F=34$ and 36 bases have been performed. If 
the difference between binding energies $B(N_F=36)$ and $B(N_F=34)$ exceeds 
2 keV, then extra calculations with $N_F=38$ have been carried out.  If full convergence
at $N_F=38$ is not reached, then extrapolation procedure of Ref.\ \cite{TOAPT.24} has 
been used  to define the solution corresponding to infinite fermionic basis. Note that 
exact solution is independent of scaling factor $f$ and deformation of basis. Thus, 
the $f=1.4$ value has been used to obtain such solutions since the binding
energies typically converge faster at this value of $f$.  Global calculations have been carried 
out for all experimentally known 882 even-even nuclei (see Ref.\ 
\cite{AME2020-second}).

\begin{turnpage}
\begin{figure*}[htb]
 \begin{center}
   \centering
   \includegraphics*[width=16.0cm]{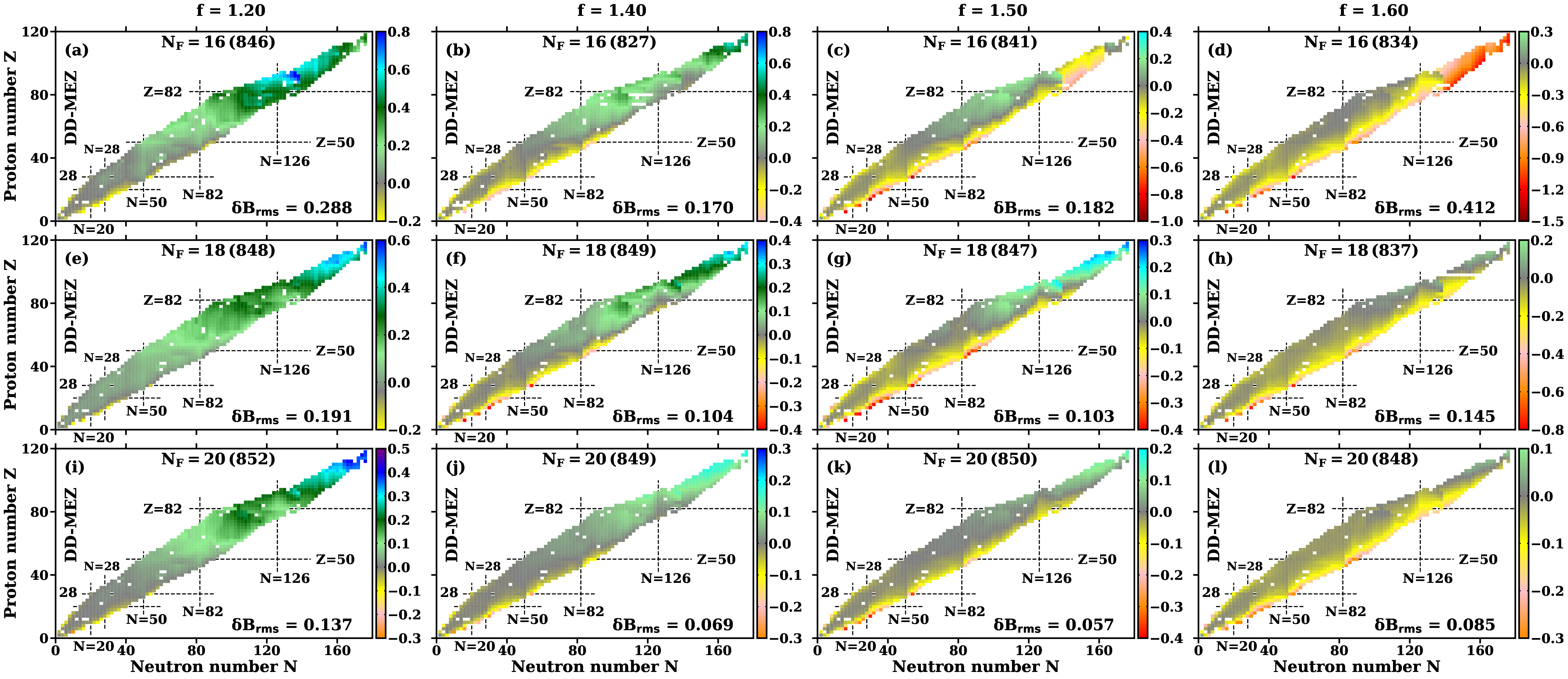}
   \caption{The same as Fig.\ \ref{global-NL5Z} but for the DD-MEZ functional.   
    \label{global-DDMEZ} 
    }
\end{center}
\end{figure*}
\end{turnpage}

   The global analysis is restricted to even-even nuclei for
which experimental binding energies are available in Atomic Mass Evaluation 
2020 (see Ref.\ \cite{AME2020-second}). Such a selection is in part due to the fact that these 
data are used in global fits of EDFs for which numerical accuracy in the calculations 
of the binding energies is of high importance (see Refs.\ \cite{TOAPT.24,OATDPDF.25}). 
In addition, it covers the part of nuclear chart in which the most of experimental and 
theoretical studies take place and which will benefit substantially from an improved 
numerical accuracy of the CDFT  calculations.  Moreover, the differences in neutron 
and proton density distributions
increase with approaching the neutron drip line and this may require the 
introduction of different oscillator frequencies for proton and neutron subsystems
which will substantially complicate the problem. 

  Figs.\ \ref{global-NL5Z} and \ref{global-DDMEZ} present such a comparison.
Let us first discuss  the results for the NL5(Z) functional. For $N_F=20$, 
the  $f=1.3$ scaling factor provides the best accuracy of $\delta B_{rms}=0.034$ MeV 
of the reproduction of the $N_F=\infty$ results. Moving away from these $f$ values leads 
to a substantial reduction of the 
accuracy of the reproduction of the $N_F=\infty$ results: $\delta B_{rms}$
becomes equal  to 0.157 and 0.126 MeV for $f=1.0$ and 1.6, respectively.
The $\delta B_{rms}$ values increase with decreasing $N_F$ but the rate of the 
increase depends on scaling factor. 
 The lowest change is seen for the $f=1.3$ scaling  factor.
Moving away from these $f$ values triggers the increase of the rate of the 
change of $\delta B_{rms}$ with decreasing $N_F$.

 The same features exists  also for the DD-MEZ functional (see Fig.\ 
\ref{global-DDMEZ}).  The $f=1.5$ factor provides the 
best accuracy ($\delta B_{rms}=0.057$ MeV) of 
the reproduction of the $N_F=\infty$ results at $N_F=20$ 
and slow rate of increase of $\delta B_{rms}$ with 
decreasing $N_F$. However, the $f=1.4$ factor provides comparable
but slightly worst results as compared with the $f=1.5$ one at
$N_F=18$ and 20 but slightly outperforms it at
$N_F=16$.

\begin{figure}[htb]
 \begin{center}
   \centering
   \includegraphics*[width=8.6cm]{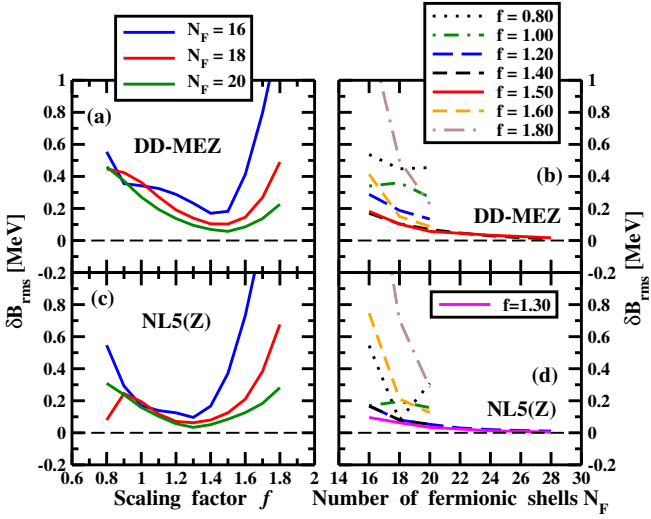}   
   \caption{Global rms  differences  $\delta B_{rms}$  between exact and truncated 
solutions as a function of scaling factor $f$ (left panels) and as a function of
$N_F$ (right panels) for all investigated combinations of $N_F$ and scaling factor $f$.
\label{global-summary} 
}
\end{center}
\end{figure}

    Fig.\ \ref{global-summary} provides  a summary of global rms  differences  
$\delta B_{rms}$  between exact and truncated solutions for the most of investigated 
combinations of $N_F$ and scaling factor $f$ including those which
are not shown in Figs.\ \ref{global-NL5Z} and \ref{global-DDMEZ}.
One can
see that for the DD-MEZ functional the minimum of $\delta B_{rms}$ is 
located  at $f=1.5$ for $N_F=20$ and at $f=1.4$ for
$N_F=16$ with both factors providing comparable accuracy at $N_F=18$
[see Fig.\ \ref{global-summary}(a)].
In the case of the NL5(Z) functional, this minimum is seen at $f=1.3$
for $N_F=16$, 18 and 20 [see Fig.\ \ref{global-summary}(b)].
For both functionals, the use of above mentioned scaling factors
instead of commonly used $f=1.0$ improves the global
rms  differences  $\delta B_{rms}$  between exact and truncated (at $N_F=20$) 
solutions by a factor of $\approx 4.6$  and $\approx 4.8$ for the NL5(Z) and DD-MEZ 
functionals, respectively.

  Figs.\ \ref{global-summary}(b) and (c) also display additional results for $\delta B_{rms}$ 
obtained at $N_F=22$, 24, 26  and 28 with the $f=1.4$ and 1.5 scaling factors
for DD-MEZ and $f=1.2$ and 1.3 ones for NL5(Z). One can see that 
for both functionals global rms  differences between exact and truncated solutions almost 
linearly decrease with increasing $N_F$.  This provides a useful tool in the selection of
the basis size which generates required global accuracy of the description of exact solutions.

\begin{figure}[htb]
 \begin{center}
   \centering
   \includegraphics*[width=8.6cm]{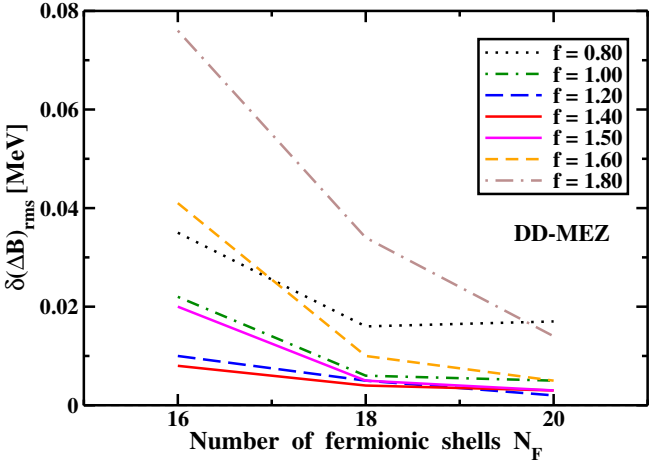}   
   \caption{Global rms deviations $\delta(\Delta B)_{rms}$ of binding energies 
   due to the deformation $\beta_0$ of the basis as a function of $N_F$ for different values
   of scaling factor $f$.    
   \label{global-deformation} 
}
\end{center}
\end{figure}

 In deformed calculations, one can also use the deformation of basis $\beta_0$ as  
a variational parameter\footnote{ In triaxial nuclei, the $\gamma_0$-deformation of the 
basis is an additional parameter which defines the HO basis (see Appendix in Ref.\  
\cite{AKR.96}). To our knowledge, it is only in this reference that an attempt to use 
the parameters of the HO basis as variational ones in the calculations of deformed 
nuclei has been undertaken in the CDFT framework. However, it was concluded that 
this does not improve the situation and that it is better to use large fermionic basis which 
drastically reduces the need  for optimization of $\hbar \omega_0$,  
$\beta_0$ and $\gamma_0$ as well as the  deviation from an exact solution. 
The present analysis of the dependence of binding energies on the quadrupole
deformation $\beta_0$ of the basis (see discussion of Fig.\ \ref{global-deformation} below) suggest 
that for large $N_F$ values such dependence is weak. It is reasonable to expect weak
dependence of binding energies on $\beta_0$ and $\gamma_0$ of the basis in 
triaxial nuclei. Thus, the optimal values of $\hbar \omega_0$ defined from the
present study of spherical and axially deformed nuclei should be also applicable
to triaxial ones.
}. For example, simultaneous variation of $\hbar \omega_0$ 
and  $\beta_0$ has been used in the DFT calculations with the Gogny force in Ref.\ 
\cite{DGLGHPPB.10}. For many years it is accepted in the CDFT community that 
the deformation of the basis close to the expected deformation of the nucleus provides 
a reasonable accuracy of the description of binding  energy in truncated calculations. 
However, it is important to evaluate a potential error due to such an approximation. In our 
global calculations four deformations of basis $\beta_0=-0.2$, 0.0, 0.2 and 0.4  are used 
to ensure the convergence to the global minimum (see Ref.\ \cite{TOAPT.24} for more
details):  among obtained solutions the lowest in energy solution is assigned to a global 
minimum.  In many cases, the solutions with different values of deformation of the basis 
converge to the same global minimum.  This allows to evaluate the impact of the deformation of 
the basis on binding energies by considering global rms deviations $\delta(\Delta B)_{rms}$ 
related to the  $\Delta B(Z,N)=B_{max}(Z,N) - B_{min}(Z,N)$ quantity.  This quantity compares 
the maximum and minimum binding energies in a global minimum of a given nucleus obtained 
with at least two out of four indicated above deformations of the basis. This quantity provides 
an estimate on how much the binding energy can be modified by the variation of the deformation 
of the basis in a reasonable interval  in a given truncated basis. In reality the calculation
error due to the selection of the deformation of basis is lower than  this estimate since we 
always select the lowest in energy solution amongst those obtained with four values of the deformation
of the basis.

  Fig.\ \ref{global-deformation} compares the $\delta(\Delta B)_{rms}$ values calculated 
for different combinations of $f$ and $N_F$ for the DD-MEZ functional. Similar
results are also obtained for the NL5(Z) CEDF and thus they are not shown. One
can see that the best and comparable results exist for $f=1.2$ and 1.4: the 
$\delta(\Delta B)_{rms}$ values are below 5 keV for $N_F=18$ and 20. Note that 
global rms deviations $\delta(\Delta B)_{rms}$ increase on moving away from these $f$ 
values. Thus, one can conclude that the optimization of the HO basis substantially reduces 
the dependence of binding energies on the deformation of the basis. 

   Fig.\ \ref{global-deformation} also shows that in general for a given scaling factor $f$  
the $\delta(\Delta B)_{rms}$ values decrease with increasing $N_F$. Moreover, these 
values are smaller than the differences in binding energies caused by the use of different 
scaling factors (see Figs.\ \ref{global-NL5Z} and  \ref{global-DDMEZ}). This feature together 
with the fact that the calculation error due to  the selection of the deformation of basis is lower 
than  the $\delta(\Delta B)_{rms}$ values explains why no attempt to optimize the deformation 
of basis has been undertaken in the present study.

\section{ General discussion on the optimization of the HO basis 
               for meson exchange functionals }
\label{HO-ME-opt}

Based on the results obtained in the present paper one can make the conclusions
on the range of the scaling factors $f$ and the $N_F$ values which, in general, are 
suitable for the CDFT calculations in moderately sized fermionic basis.
Fig.\ \ref{dep-on-f} shows that for all classes of the functionals there is a rapid variation 
of binding energies as a function of scaling factor $f$ for $f\leq 1.0$ 
in the $N_F=16$ and 20 bases which are frequently used in modern calculations.
This feature somewhat depends on the nuclei and it can lead to unexpected biases which, 
for example, are clearly visible in Fig.\ \ref{global-NL5Z}(a): the truncated $(N_F=16, f=1.0)$ 
solutions are more bound than the exact  ones for sublead region but then the situation sharply reverses 
for lead region and actinide nuclei\footnote{Note that this feature becomes even more
pronounced for the $(N_F=16, f=0.8)$ solutions not shown in Fig.\ \ref{global-NL5Z}.}. The 
increase of the $N_F$ value to 18 removes this bias [see Fig.\ \ref{global-NL5Z}(e)]  since the 
region of rapid changes of binding energies with variation of $f$ moves to lower $f$ values
with increasing $N_F$ (see Fig.\ \ref{dep-on-f}). 

Quite good global agreement between truncated and exact calculations is obtained for the 
$(N_F=18, f=0.8)$ case in the NL5(Z) functional [see Fig.\ \ref{global-summary}(c)]. However,
such combination is not recommended since (i) moderate changes of $f$ trigger quite large 
changes in  $\delta B_{rms}$  and (ii) there is a large staggering of the $\delta B_{rms}$
values for $f=0.8$ as a function of $N_F$. The latter feature is in contrast to a general 
decreasing trend of $\delta B_{rms}$ with increasing $N_F$ seen for other values of
$f$ (see Fig.\ \ref{global-summary}).

    The combination of high $f$  and low $N_F$ values can lead to a similar sharp transition 
which is seen around $N=136$ in the  $(N_F=16, f=1.6)$ calculations with the NL5(Z) and DD-MEZ 
functionals [see Figs.\ \ref{global-NL5Z}(d) and \ref{global-DDMEZ}(c)].
However, in that case this sharp transition is caused by the fact that  the $N_F^{trans}$ value at 
which the pattern A convergence curve changes from strongly downsloping  to slowly approaching 
the exact value depends both on the $f$ value and the nucleus (see Fig.\ \ref{conver-208Pb-240Pu}). 
For a given scaling factor $f$, the $N_F^{trans}$ value decreases with decreasing the mass of
nucleus. Thus, in the part of nuclear chart below $N\approx 136$ above mentioned truncated 
calculations correspond to a part of convergence curve which is slowly approaching the exact solution. 
This explains small differences between the exact and truncated calculations  [see Figs.\ \ref{global-NL5Z}(d) 
and \ref{global-DDMEZ}(c)]. However, above this neutron number the difference between such
calculations drastically increases since truncated calculations are carried out on strongly downsloping
part of convergence curve. 

      By selecting  $f\approx 1.3$ ($f\approx 1.5$) in the  NLME (DDME) functionals one 
guarantees that the calculations for all nuclei of interest are carried out on the part of convergence curve 
which slowly approaches the exact solution and which starts at  $N_F^{trans}$ value which is the lowest 
among  considered values of $f$ (see Fig.\ \ref{conver-208Pb-240Pu}).  In addition, this selection ensures that the 
$f$ value is located in the region of moderate changes of binding energies as a function of scaling 
factor $f$ (see Fig.\ \ref{dep-on-f}). Moreover, it  guarantees that no above mentioned 
numerical biases appear and that quite accurate description is obtained even in relatively small
$N_F=16$ basis which is characterized by a rather accurate ($\delta B_{rms} = 0.096$ MeV
for NL5(Z) and $\delta B_{rms} = 0.170$ MeV for DD-MEZ)
reproduction of exact results  [see Fig.\ \ref{global-NL5Z}(c) and
Fig.\ \ref{global-DDMEZ}(b)].  
 
   The present analysis suggests that global optimizations of the HO basis  can also
improve the performance of the DFTs based on the Skyrme and Gogny functionals. At
present, such global studies are not available and the selection of the harmonic oscillator 
frequencies is based either on a very limited set of data, input from other models (see 
Ref.\ \cite{DD.97-HFODD}) or on simplified arguments (see Ref.\ \cite{D2}).
For example, the HO frequency 
$\hbar \omega_0 = 1.2 \times 41 A^{-1/3}$ is used in many Skyrme  DFTs calculations (see 
Refs.\ \cite{DD.97-HFODD,UNEDF0}).  Another example are Refs.\ \cite{D2,PH.17} in
which the $\hbar \omega_0$ value of the Gogny forces is defined from the charge radii 
of $^{16}$O and $^{90}$Zr using restricted Hartree-Fock approximation. However, this 
contradicts  to the fact that in sufficiently large fermonic basis the charge radii are 
independent of oscillator frequency $\hbar \omega_0$ for a large range of scaling
factor $f$ (see Fig.\ \ref{bos-fer-radii}).

\section{Mass dependence of scaling factors $f$}
\label{Mass-dep}

\begin{figure}[htb]
 \begin{center}
   \centering
   \includegraphics[width=8.6cm]{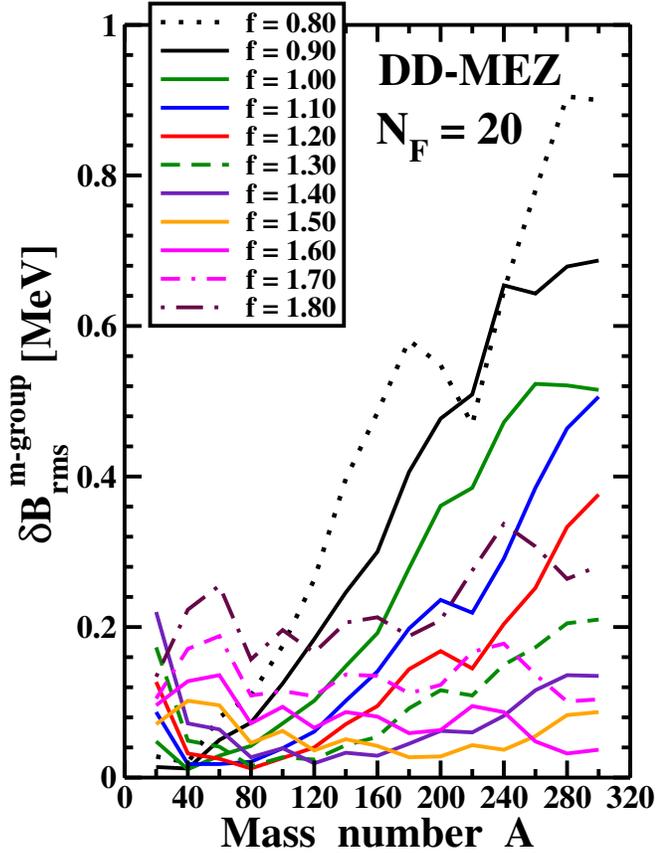}  
   \caption{The rms differences $\delta B_{rms}^{m-group}$ as a function
   of mass number $A$ for different values of scaling factor $f$. For each 
   $m$-group  the results are given at $A=20 m$. See text for further details.
  \label{f-evolution} 
}
\end{center}
\end{figure}

\begin{figure}[htb]
 \begin{center}
   \centering
   \includegraphics*[width=8.6cm]{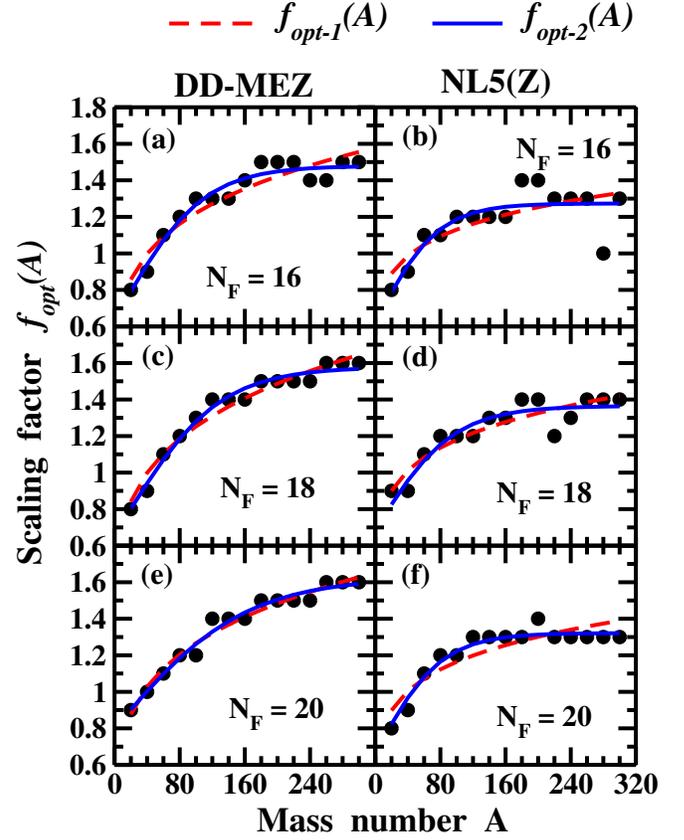}  
   \caption{The functions  $f_{opt-1}(A)$ and $f_{opt-2}(A)$ fitted to the
   $f_{opt}(m)$ data shown by black solid circles. The optimized parameters
   of these functions are provided in Table \ref{Table-opt-par}. Note that in the
   case of $N_F=16$ and NL5(Z) functional (panel(b)), the outlier at ($f=1.0, A=280$)
   is excluded from fitting procedure. 
   \label{f-mass-dep} 
}
\end{center}
\end{figure}

      \begin{table*}
	\centering
	\caption{The dependence of $\delta B_{rms}$ and $\delta (r_{ch})_{rms}$	
	on scaling factor $f$ and $N_F$ used in global calculations. $\delta (r_{ch})_{rms}$ 
	is global rms difference between the values of charge radii obtained in the $(N_F, f)$ calculations 
	and those defined in the $N_F=\infty$ ones. Globally fixed values of  $f_{opt}=1.5$ and $f_{opt}=1.3$ 
	are used for the DD-MEZ and NL5(Z) functionals, respectively. See text for further details.
\label{Table-compar}	
	}
	\begin{tabular}{cccccccccc}
	\hline \hline 
	CEDF & $N_F$ & \multicolumn{4}{c|}{$\delta B_{rms}$ [MeV]} & \multicolumn{4}{c|}{$\delta (r_{ch})_{rms}$ [fm]}\\ 
	\cline{3-10} 
	     &   & $f = 1.0$ & $f_{opt}$ & $f_{opt}(m)$   & $f_{opt}(A)$   & $f = 1.0$ & $f_{opt}$  &  $f_{opt}(m)$   & $f_{opt}(A)$  \\
	\hline 
	DD-MEZ & 16 & 0.342 & 0.182 &  0.111 & 0.114  & 0.00126 & 0.00194 &  0.00142 & 0.00148 \\ 
	       & 18 & 0.363 & 0.103 &  0.057 & 0.060  & 0.00107 & 0.00104 &  0.00062 & 0.00056 \\
	       & 20 & 0.271 & 0.057 &  0.029 & 0.031  & 0.00083 & 0.00092 &  0.00037 & 0.00030 \\
	\hline 
	NL5(Z) & 16 & 0.170 & 0.096 & 0.061 & 0.069 & 0.00157 & 0.00135 &  0.00116 & 0.00113 \\ 
	       & 18 & 0.195 & 0.063 & 0.038 & 0.041 & 0.00114 & 0.00085 &  0.00051 & 0.00055 \\
	       & 20 & 0.157 & 0.034 & 0.023 & 0.025 & 0.00085 & 0.00058 &  0.00032 & 0.00031 \\
	\hline \hline  
	\end{tabular} 
\end{table*}

  There are considerable variations in the pattern of density distribution in nuclei:
these densities  are narrow in radial coordinate in very light nuclei but with increasing  
mass number  the pattern typical for the Fermi distribution develops with the region 
of near  constant density extending to a large radial coordinate (see Fig.\ 2.4 in Ref.\ 
\cite{NilRag-book}).  Thus, it is important to understand whether there are some 
correlations between the optimal value of scaling factor $f$ and the evolution of the
density of nuclei across the nuclear chart.

     \begin{table}
	\centering
	\caption{ The parameters of the functions $f_{opt-1}(A)$ and $f_{opt-2}(A)$ fitted
	to the $f_{opt}(m)$ data shown by solid circles in Fig.\ \ref{f-mass-dep}.
	These parameters are defined for the DD-MEZ and NL5(Z) functionals.	
	The $\delta_{rms}$ values provide the information on the quality of the
	fit.
\label{Table-opt-par}	
}
	\begin{tabular}{ccccccccc}
	\hline 
	CEDF & $N_F$ & \multicolumn{3}{c}{
	$f_{opt-1}(A)$}  & \multicolumn{4}{c}{$f_{opt-2}(A)$}  \\ 
	\cline{3-9} 
	     &   & c & $\alpha$ & $\delta_{rms}$ & c & $\alpha$ & $A_0$ & $\delta_{rms}$ \\
	\hline \hline
	DD-MEZ & 16 & 0.443 & 0.220 & 0.066 & 1.479 & 0.021 & 13.3 & 0.045 \\ 
	       & 18 & 0.392 & 0.252 & 0.046 & 1.578 & 0.018 & 18.2 & 0.034 \\
	       & 20 & 0.443 & 0.228 & 0.037 & 1.626 & 0.013 &  2.8 & 0.033 \\
	\hline 
	NL5(Z) & 16 & 0.487 & 0.186 & 0.065 & 1.330 & 0.020 & -1.06  & 0.054 \\ 
	       & 18 & 0.544 & 0.168 & 0.065 & 1.373 & 0.019 & -10.7  & 0.059 \\
	       & 20 & 0.561 & 0.157 & 0.068 & 1.307 & 0.030 &   7.8  & 0.022 \\
	\hline 
	\end{tabular} 
\end{table}
 
 To reveal such correlations the results presented in Figs.\ \ref{global-NL5Z} and \ref{global-DDMEZ} 
are rearranged into the groups containing the nuclei with mass numbers between 
$2 + 20 * (m-1)$ and $20 + 20 * (m-1)$ where $m=1, 2, 3, ...$.  Then for each $m$-group 
the rms  difference  $\delta B_{rms}^{m-group}$ between the binding energies obtained 
in infinite and truncated bases is defined for different values of  scaling 
factor $f$.  Fig.\ \ref{f-evolution} shows the results of such procedure for $N_F=20$ 
and DD-MEZ functional but similar results are obtained also for $N_F=16$ and 18 
and the NL5(Z) functional. One can see that for $m=1$ ($A=2-20$) group the best 
reproduction of exact results  is obtained with $f=0.9$ but with increasing mass number 
the quality of the description  of exact results by this $f$ value deteriorates rapidly so that  
$\delta B_{rms}^{m-group}\approx 0.7$ MeV for the $A=262-280$ group. With increasing
$A$ the $f$ value which provides the best description of the exact results gradually increases.
For example, in the $N_F=20$ calculations with the DD-MEZ functional the best description of 
exact results in the $A=22-40$, $A=42-60$, $A=62-100$,  $A=102-160$, $A=162-240$ and 
$A=242-300$ mass regions is provided by scaling factors 1.0, 1.1, 1.2, 1.4, 1.5 and 1.6, 
respectively (see Fig.\ \ref{f-evolution}). 

   Such scaling factors, which provide the best description of exact results in 
a given truncated basis, are labelled as optimal scaling factors $f_{opt}(m)$ for 
a given $m$-group (or a given mass range).  They are shown as solid circles in Fig.\ 
\ref{f-mass-dep}.  One can see that, in general,  they increase with increasing
mass. However, in the $N_F=18$ and 20 calculations the $f_{opt}(m)$ values the 
NL5(Z) functional saturate for  $A>160$ values  but it is not clear whether
such saturation is reached in the DD-MEZ functional.  The use of $f_{opt}(m)$ leads 
to a substantial (by a  factor of $\approx 1.5-2.0$) improvement  in  global rms 
differences  $\delta B_{rms}$  between exact results and those obtained  in  truncated 
basis as compared with  the $\delta B_{rms}$ values generated with the  $f_{opt}$ 
value fixed across the nuclear  chart (see Table \ref{Table-compar}). Such an improvement
is substantially larger (by a factor ranging from $\approx  3$ to $\approx 9$ dependent 
on functional and $N_F$) when the $\delta B_{rms}$ values obtained with $f_{opt}(m)$ and 
$f=1.0$ are compared in Table \ref{Table-compar}.
 
\begin{figure*}[htb]
 \begin{center}
   \centering
   \includegraphics*[width=14.0cm]{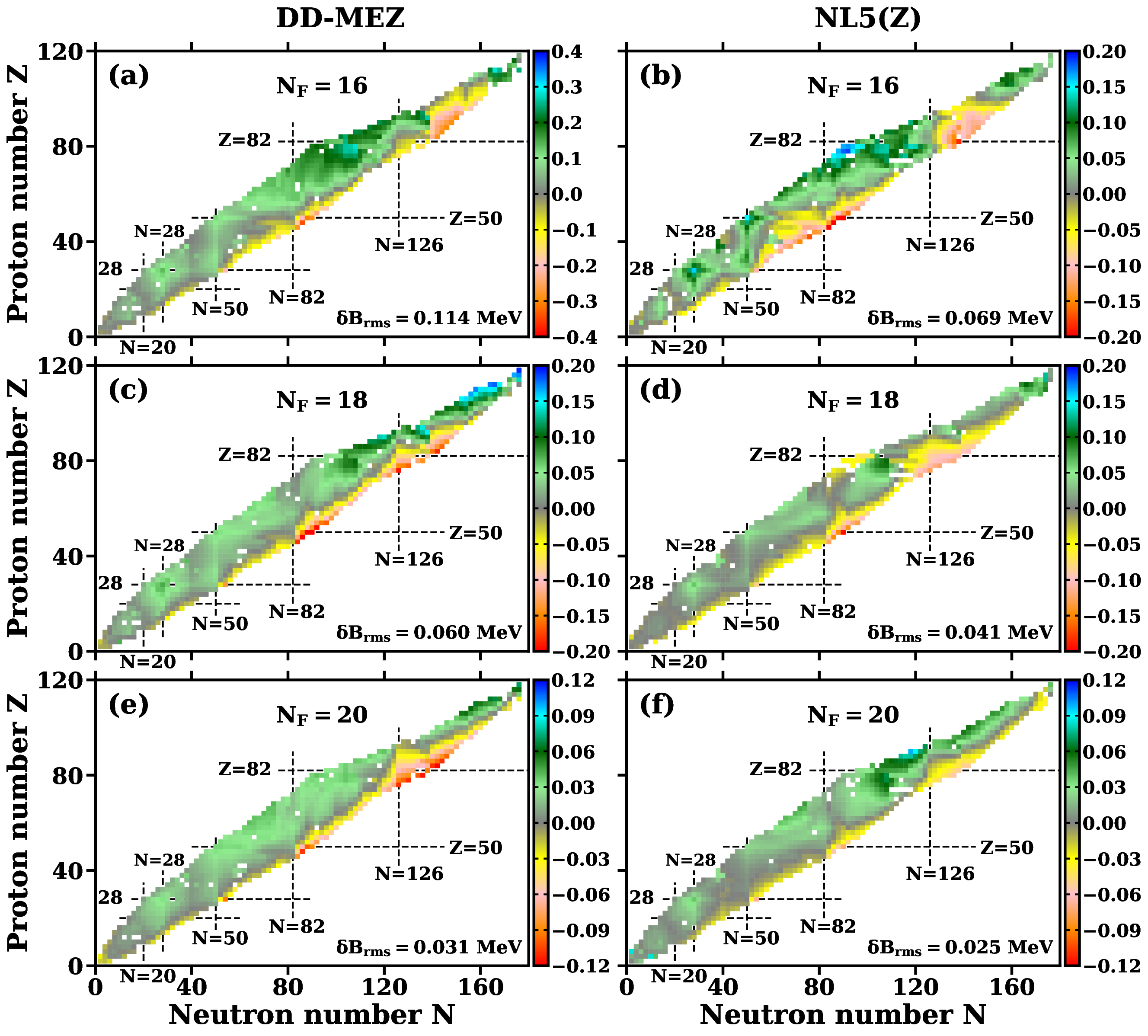}  
   \caption{The same as Fig.\ \ref{global-NL5Z}  but for the results obtained 
with mass dependent scaling factor $f_{opt-2}(A)$ defined by Eq.\ (\ref{eq-2}) 
and  parameters of Table \ref{Table-opt-par}. The results are presented for the DD-MEZ 
and NL5(Z) functionals.        
   \label{global-smooth-mass-dep} 
}
\end{center}
\end{figure*}

   The $f_{opt}(m)$ is a step function and its use is recommended for the calculation of 
individual nuclei.
However,  the transition from one $m$-group to another 
characterized by different  $f_{opt}(m)$ value creates  a numerical step in binding energies at 
the boundary of these  $m$-groups in the calculations with truncated basis.
 This step affects also two-particle separation 
energies. Thus, for global calculations a smooth dependence of  scaling factor $f$ on 
mass is required to avoid the appearance of such a step. Two approximate functions 
(power and sigmoid)  
\begin{eqnarray}
	f_{opt-1}(A) &=& cA^\alpha, \label{eq-1} \\
	f_{opt-2}(A) &=&  \frac{c}{1+e^{-\alpha(A-A_0)}},  \label{eq-2}  
\label{fit-eq} 
\end{eqnarray} 
were used for the definition of smooth mass dependence of scaling factor $f$. 
%
%
It turns out that the sigmoid function $f_{opt-2}(A)$ 
provides a better quality fit to $f_{opt}(m)$ (see Fig.\ \ref{f-mass-dep} and  Table 
\ref{Table-opt-par}). Thus only it is used  in further analysis.

Note that the "smoothness" of the $f_{opt}(m)$ function as a function of 
$A$ and thus the  quality of its  description by the $f_{opt-1}(A)$ and $f_{opt-2}(A)$ ones 
improves with increasing $N_F$  (see Fig.\ \ref{f-mass-dep} and  Table \ref{Table-opt-par}).
    For a given functional, the spread of  the $f_{opt-2}(A)$ functions obtained with $N_F=16$,
18 and 20 at given $A$ is typically  below 0.1 (see Fig.\ \ref{f-mass-dep}). This points to a 
reasonable stability of mass dependent  $f_{opt-2}(A)$ function on the choice of $N_F$
and its applicability for other functionals in a given class (NLME or DDME) of the functionals.

   Fig.\ \ref{global-smooth-mass-dep} shows the accuracy of the description of the 
$N_F = \infty$ results when the scaling factor $f$ is provided by the $f_{opt-2}(A)$ function.
One can see that these results are substantially better than those obtained with globally 
fixed $f_{opt}=1.5$ (DD-MEZ) and $f_{opt}=1.3$ [NL5(Z)] scaling factors (compare Fig.\ 
\ref{global-smooth-mass-dep} with Figs.\ \ref{global-NL5Z} and \ref{global-DDMEZ} and 
see Table \ref{Table-compar}). They are slightly worst as compared with those 
provided by the $f_{opt}(m)$ step function (see  Table \ref{Table-compar}) but the
issue of numerical step at the boundary of adjacent $m$-regions with different
values of $f_{opt}(m)$ is avoided. There are still some unresolved trends in the
isospin direction the importance of which decreases with increasing $N_F$ (see
Fig.\ \ref{global-smooth-mass-dep}). They can probably be addressed by the use of 
different scaling factors $f$ for proton and neutron subsystems but such a study is 
beyond the scope of the present paper.

   Table \ref{Table-compar} also provides the information on the accuracy of the
description of charge radii in the $(N_F,f)$ schemes as compared with the
$N_F=\infty$ results. This accuracy is better than the one
obtained in experiment (see Ref.\ \cite{AM.13}) for all considered  
$(N_F,f)$ schemes.  The $\delta (r_{ch})_{rms}$ values
decrease with increasing $N_F$ and, in general, with optimization
of scaling factor $f$ (compare the $f_{opt}$,  $f_{opt}(m)$,   $f_{opt}(n)$ results
with the $f=1.0$ ones in Table \ref{Table-compar}).  It is also significantly better
than the global accuracy of the description of experimental charge radii
$\Delta(r_{ch})_{rms} \approx 0.025$ fm obtained in the CDFT
calculations (see Refs.\ \cite{TA.23,OATDPDF.25}).

\section{Conclusions}
\label{Concl}
 
The main goal of the present study is further development of covariant 
density functional theory towards more accurate description of binding 
energies across the nuclear chart within moderately sized fermionic basis.
This is achieved both by a better understanding  of the convergence of these 
energies as a function of the size of the basis and by a global optimization 
of harmonic oscillator frequency of the basis.
   
   The main results  can be summarized as follows.
\begin{itemize}
\item
  Two basis  (fermionic plus bosonic) structure of the CDFT for meson exchange 
functionals  is unique in nuclear physics. In asymptotic (monotonic) part
of the convergence curve, the fermionic and mesonic (bosonic) energies converge to an
exact solution from above and below with increasing the size of fermionic basis, 
respectively. The balance of the rates of 
the convergence of these energies depends on oscillator frequency 
$\hbar \omega_0= f\times 41 A^{-1/3}$ MeV.  As a result, the total binding energies 
for the ME functionals can converge to exact (infinite basis) solution either from 
below or from above dependent on  scaling factor $f$. This allows 
to define the optimal value of $f$  which provides the best reproduction of infinite 
basis results starting from relatively low value of $N_F$. In contrast, point 
coupling functionals do not contain mesons and as a consequence their total
binding energies in the asymptotic part always converge to the exact solution 
from above. This is similar to many non-relativistic theories.

\item
  Based on global studies of the dependence of binding energies on scaling 
factor  $f$ and the number of fermionic shells $N_F$ benchmarked with 
respect of infinite basis solutions the optimal $f$ values have been defined
for the ME functionals. They are $f=1.3$ and $f=1.5$ for  the 
NL5(Z) and  DD-MEZ functionals when scaling factor $f$ is globally
fixed.  These $f$ values  provide very high accuracy of the
calculations in moderately sized $N_F=20$ basis:  global rms differences  
$\delta B_{rms}$ between exact and truncated solutions are  only
$0.034$ MeV and $0.057$ MeV  for the 
NL5(Z) and DD-MEZ functionals, respectively. They are by a factor 
of  $\approx 4.6$ and $\approx 4.8$ better  than those obtained with traditionally
used $f=1.0$ scaling factor.  The introduction of mass dependence 
of scaling factors $f_{opt-2}(A)$ via Eq.\ (\ref{eq-2}) and the parameters defined in 
Table \ref{Table-opt-par} leads  to a further improvement of the accuracy and reduces 
the $\delta B_{rms}$  values down to  $0.025$ MeV and $0.031$ MeV for above 
mentioned  functionals.

\item 
   There are very strong correlations between optimal values of $f$ obtained
in global calculations and those which follow from the analysis of selected
set of spherical and deformed nuclei such as $^{48}$Ca, $^{208}$Pb, 
$^{240}$Pu and $^{304}$120 (compare, for example,  Figs.\ \ref{global-DDMEZ} 
and \ref{global-summary} with Fig.\ \ref{conver-208Pb-240Pu}). 
Using this fact and detailed analysis of such nuclei with different functionals it 
was concluded that above discussed globally fixed ($f_{opt}$) and mass 
dependent ($f_{opt-2}(A)$) scaling factors defined for the NL5(Z) and 
DD-MEZ functionals are also optimal ones for the NLME and DDME classes 
of the functionals.  The analysis of the 
$^{48}$Ca and $^{208}$Pb nuclei indicates that the best convergence of 
binding energies is obtained in the PC functionals  for the values of $f$ 
ranging from 1.2 up to  1.6 in  light nuclei and ranging from 1.4  up to 1.8 
in heavy ones.

\end{itemize}

\section{ACKNOWLEDGMENTS}

 This material is based upon work supported by the U.S. Department of Energy,  
Office of Science, Office of Nuclear Physics under Award No. DE-SC0013037.

\bibliography{references-46-optimization-HO-basis.bib}

\end{document}